\documentclass{jfm}
\usepackage{amssymb}
\usepackage{amsmath}
\usepackage{graphicx}
\usepackage{dcolumn}
\usepackage{bm}
\usepackage{epsfig}
\usepackage{float}
\usepackage[T1]{fontenc}
\usepackage{ae,aecompl}
\usepackage{epstopdf, epsfig}
\usepackage{color}
\usepackage{appendix}
\usepackage[export]{adjustbox}
\usepackage{graphicx}
\usepackage{epstopdf, epsfig}
\usepackage[section]{placeins}
\usepackage[latin1]{inputenc}
\usepackage{subfiles}
\usepackage{natbib}
\usepackage{hyperref}
\hypersetup{
    colorlinks = true,
    urlcolor   = red,
    linkcolor = blue,
    citecolor = blue,
    filecolor = black,
}
\allowdisplaybreaks[4]
\shorttitle{DB multi-scale modeling of non-equilibrium multiphase flows}
\shortauthor{Gan, Xu, Lai, Li, Sun, and Succi}
\affiliation{\aff{1}Hebei Key Laboratory of Trans-Media Aerial Underwater Vehicle, School of Liberal Arts and Sciences, North
China Institute of Aerospace Engineering, Langfang 065000, China
\aff{2}Laboratory of Computational Physics, Institute of Applied Physics and
Computational Mathematics, P. O. Box 8009-26, Beijing 100088, China
\aff{3}State Key Laboratory of Explosion Science and Technology, Beijing
Institute of Technology, Beijing 100081, China
\aff{4}HEDPS, Center for Applied Physics and Technology, and College of
Engineering, Peking University, Beijing 100871, China
\aff{5}College of Mathematics and Statistics, FJKLMAA, Center for Applied
Mathematics of Fujian Province (FJNU), Fujian Normal University, Fuzhou
350117, China
\aff{6}School of Physics, Beijing Institute of Technology, Beijing 100081, China
\aff{7}Center for Life Nano Science at La Sapienza, Fondazione Istituto Italiano
di Tecnologia, Viale Regina Margherita 295, 00161, Roma, Italy
\aff{8}Physics Department and Institute for Applied Computational Science, John
A. Paulson School of Applied Science and Engineering, Harvard University,
Oxford Street 29, Cambridge, MA 02138, USA}

\begin{document}
\title{Discrete Boltzmann multi-scale modeling of non-equilibrium multiphase flows}
\author{Yanbiao Gan\aff{1}, Aiguo Xu\aff{2,3,4} \corresp{\email{Xu%
\_Aiguo@iapcm.ac.cn}}, Huilin Lai\aff{5}, Wei Li\aff{1}, Guanglan Sun\aff{1,6}, \and %
Sauro Succi\aff{7,8}}
\maketitle

\begin{abstract}
The aim of this paper is twofold: the first is to formulate and validate a multi-scale discrete Boltzmann method (DBM) based on density functional kinetic theory for thermal multiphase flow systems, ranging from continuum to transition flow regime; the second is to present some new insights into the thermo-hydrodynamic non-equilibrium (THNE) effects in the phase separation process.
Methodologically, DBM includes three main pillars:
(i) the determination of the fewest kinetic moment relations, which are required by the description of significant THNE effects beyond the realm of continuum fluid mechanics, (ii) the construction of appropriate discrete equilibrium distribution function recovering all the desired kinetic moments, (iii) the detection, description, presentation and analysis of THNE based on the moments of the non-equilibrium distribution ($f-f^{(eq)}$).
The incorporation of appropriate additional higher-order thermodynamic kinetic moments considerably extends the DBM's capability of handling larger values of the liquid-vapor density ratio, curbing spurious currents, and ensuring mass-momentum-energy conservation.
Compared with the DBM with only first-order THNE \citep{2015-Gan-SM-PS}, the model retrieves kinetic moments beyond the third-order super-Burnett level, and is accurate for weak, moderate, and strong THNE cases even when the local Knudsen number exceeds $1/3$.
Physically, the ending point of the linear relation between THNE and
the concerned physical parameter
provides a distinct criterion to identify whether the system is near or far from equilibrium. Besides, the surface tension refrains the local THNE around the interface, but expands the THNE range and strengthens the THNE intensity away from the
interface through interface smoothing and widening.
\end{abstract}

\section{Introduction}\label{sec1}
Multiphase flows with phase transitions and heat transfer are ubiquitous in
natural and industrial processes, such as atmospheric phenomena,
material and food processing, petrochemical engineering, bio-medical as well as  life
science~\citep{Multiphase-Book,ZDY-PhysRep-2019,Succi-RevModPhys-2019,WB-2021-JFM}.
Therefore, establishing accurate, reliable and efficient models and computational strategies for predicting their flow behavior,
will deepen our understanding of the fundamental and underlying physical
mechanisms behind multiphase flows.
Besides major academic significance as complex phenomena far from equilibrium,
they also bear essential industrial value.

Nevertheless, multi-scale modeling and simulation of such a
complex system is a long-standing challenge~\citep{Luo-2009-JMM,ZDY-PhysRep-2019,Succi-RevModPhys-2019}.
The challenge arises from the common and unique features of these fluids:
(i) complex multi-scale structures and their cross-scale correlations, such as particles, bubbles,
droplets and clusters; (ii) rich and abundant evolving interfaces, such as
material and mechanical interfaces; (iii) complex forces, relaxations and
responses, such as gradient forces, interparticle forces, external forces, and
the nonlinear coupling among them; (iv) competition between various spatiotemporal
scales and kinetic modes, such as growth, deformation, breakup, cavitating, boiling, and even turbulence.

In general, a multiphase flow system is in a global or local
thermo-hydrodynamic non-equilibrium (THNE) state with fluctuations~\citep{2020-Wagner-PRL}, including significant hydrodynamic
and thermodynamic non-equilibrium  (HNE and TNE, respectively) effects which
may undermine the validity of macroscopic models, such as the most commonly used Navier-Stokes (NS) equations. Simplifying assumptions, such as thermal equilibrium across a
liquid-vapor interface during phase separation can lead to
erroneous predictions~\citep{2016-Chemical-Review}.
A possible solution for accessing detailed THNE effects
is to use particle-based methods, for example, molecular dynamics or direct
simulation Monte Carlo method~\citep{Kang-PRE-2017,Kang-PRE-2020}. However,
the spatiotemporal scales and geometries that the two schemes can afford are
extremely small and highly idealized compared with those of practical applications.

As a mesoscopic approach and a natural bridge connecting microscopic and
macroscopic models, suitably extended versions of the Boltzmann equation
can in principle describe the complex non-equilibrium thermo-hydrodynamics for the full
spectrum of flow regimes~\citep{Chapman-Book}.
However, the nonlinearity, multidimensionality, and integro-differential nature of the
collision term pose a formidable challenge to its direct solution.
The difficulty in using the original Boltzmann equation
has prompted the development of approximate and simplified kinetic models
that relinquish much of its computational complexity while preserving its most relevant physics in point. The versatile lattice Boltzmann method (LBM)~\citep{benzi1992lattice,chen1998lattice,Wagner1998breakdown,Succi2001Book,2018-Succi-Book,Shu-JCP-2006,
sbragaglia2007generalized,2012-Xu-FOP,2013-Shuchang-Book,Succi-RevModPhys-2019,Tian-JFM-2021,falcucci2021extreme,WYK-PRE-2022,2022CNF} and the recently developed discrete Boltzmann modeling method (DBM)~\citep{2015-Gan-SM-PS, 2016-Lai-PRE-RTI,lin2017multi,2018-Gan-PRE, 2019-Gan-KHI,
2019-Zhang-SM-PS,2020-Chen-POF-RTKH,2021-ZG-POF-RTI,2021-Xu-AAS,2021-Xu-AAAS,2021-Xu-JCP},
belong precisely to this class of modern non-equilibrium methods.

As for the LBM research, there are two complementary branches.
The first is a physics-inspired construction method, while the second is a
numerical scheme for solving partial differential equation(s), such as
wave equation~\citep{YGW-2000-JCP,Lai-2011-PRE}, convection-diffusion
equation~\citep{GZL-2009-PRE,Du-2020-AML,CZH-2021-PRE},
Poisson equation~\citep{CZH-2008-AMM}, Laplace equation~\citep{YGW-2009-AMC},
Fisher equation~\citep{Succi-Fisher-equation}, Burgers equation~\citep{Burgers-1996-SIAM-JSC,Duan-JCAM-2007, SWP-2011-CNSNS}, 
Benjamin-Ono equation~\citep{Benjamin-Ono-Lai}, Korteweg-de Vries equation~\citep{KDV-GuoBL,KDV-WHM}, Klein-Gordon-Zakharov equation~\citep{KGZ-WHM},
 Zakharov-Kuznetsov equation~\citep{ZK-WHM-2017,ZK-WHM-2020},
Ginzburg-Landau equation~\citep{YGW-2010-PRE,YGW-JSC-2014},
Kuramoto-Sivashinsky equation~\citep{Lai-2009-PA,KSE-2018-CF}, Schr\"{o}dinger equation~\citep{Succi2001Book,SCH-2020-AML}, etc.
These two branches build on similar, yet not equal, construction rules.
DBM was developed from the first branch and represents a mesoscopic
modeling method for fluid flows.

Although the original Boltzmann equation and its approximations work only
for dilute gas systems, they can be extended to multiphase flow regimes after
incorporating the non-ideal gas effects through a variety of approaches.
Actually, both LBM and DBM are particularly promising in the area of
multiphase and multicomponent flows, mainly on account of their  kinetic
nature, directly inherited from the Boltzmann equation, which facilitates the
inclusion of microscopic physical interactions
as compared to numerical methods based on continuum models.

To date, many LBMs for multiphase flows have been proposed, including
the chromodynamic model \citet{RK-model}, the pseudo-potential model~\citep{SC-1993,SC-1994}, the
free-energy model~\citep{Yeomans-1995-PRL,Yeomans-1996-PRE,XGL-2003-PRE,XGL-2004-PhysA},
the kinetic-theory-based model~\citep{He-1998-PRE,He-1999-JCP}, the forcing model~\citep{Sofonea-2004-PRE,GLS-model-PRE2007},
the phase-field model~\citep{Succi-PRE-2005}, the entropic kinetic model
~\citep{2015-PRL-Karlin,2017-POF-Succi,2018-PRL-Karlin}, the unified collision model~\citep{Luo-PTRA-2021}, and so forth.
Although proposed from different perspectives, the common point of these
models is the inclusion of interparticle interactions at the mesoscopic
scale. The interparticle interactions are the underlying engine behind  the
complex THNE features of multiphase flows. The aforementioned
models and their revised versions have been successfully applied to the
study of fundamental phenomena and mechanisms of multiphase flows in science
and engineering, ranging from
droplet evaporation~\citep{Yeomans-SM-2014,Safari-PRE-2013,Safari-PRE-2014,Zarghami-PRE-2017,Qin-JFM-2019},
to droplet deformation, breakup, splashing, and coalescence
~\citep{Wagner2003role,SHU-JCP-2015-1,SHU-JCP-2015-2,Chen-JFM-2017,Liu-JFM-2018,Liang-IJHMT-2019,
wen2017chemical,WEN-PRE-2020},
 collapsing cavitation
~\citep{2011-Chen-CMA,falcucci2013direct,2015-Gonella-POF,2018-Sofonea-PRE,
2020-Shan-Ultrasonics,2022-Shan},
acoustics levitation~\citep{ZDY-Book}, nucleate boiling
~\citep{Liqing-PECS-2016,Liqing-ATE-2018,Fei-POF-2020}, ferrofluid and
electro-hydrodynamic flows~\citep{Succi-PRE-2009,NXD-PRE-2018,Chai-POF-2019},
hydrodynamic instability
~\citep{He-POF-1999-RTI,He-2001-KHI,Lee-2013-KHI,
Lianghong-2014-RTI,Lianghong-2016-RTI,Lianghong-2016-PRE-KHI,tavares2021immiscible},
dendritic growth~\citep{Succi-PRE-2005,Rojas-JCP-2015,SDK-IJHMT-2016,SDK-IJHMT-2016-2},
heat and mass transfer in porous media~\citep{Chen-IJHMT-2015,Liu-CG-2016,HYL-IJHMT-2019,Chai-IJHMT-2016,Chai-SIMA-2019},
Rayleigh--B{\'e}nard convection~\citep{pelusi2021rayleigh},
active fluid~\citep{Cates-PRL-2008,Yeomans-NatComm-2016,Gonnella-PNAS-2019,Nergo-Gonnella-SM-2019},
isotropic turbulence~\citep{perlekar2014spinodal,milan2020sub},
solid-liquid-vapor phase transition and phase ordering under various
conditions~\citep{Yeomans-1995-PRL-SD,Gonella-1997-PRL,Gonella-1998-PRL,Cates-1999-PRL,
Sofonea-2004-PRE,Gonella-2010-PRE,Gonella-2014-PRE,2017-WangHP-Ultrasonics,Ambrus-Wagner-2019-PRE,Busuioc-Sofonea-2020-CMA, HuangRZ-PRL-2021,Shu-PRE-2021}.

Notwithstanding significant progress, the highly non-equilibrium interfacial thermo-hydrodynamics, which is strongly associated with the growth kinetics and morphological evolution of
systems, is seldom reported and still poorly understood. This
topic has presented a challenge to the important front of trans-scale
modeling and simulation of multiphase flows.

Recently, we have demonstrated that the DBM offers a novel systematic scheme
and a set of convenient and efficient tool for describing, measuring, and
analyzing simultaneously THNE effects which cannot be adequately
described by traditional hydrodynamic models~\citep{2012-Xu-FOP,2015-Gan-SM-PS,2016-Lai-PRE-RTI, 2018-Gan-PRE,2019-Zhang-SM-PS,2020-Chen-POF-RTKH,2021-ZG-POF-RTI,2022-FOP-Chen,
zhang2022discrete,Zhang-AIPAdv-2022,2022-Liu-jmse}.
DBM was developed from a branch of LBM which aims to describe the non-equilibrium flows from a more fundamental level and is no longer based on the simple ``propagation $+$ collision" lattice gas evolution scenario.
The key point of DBM is to ensure that the properties to be studied do not change with model simplification and use the non-conservative kinetic moments of $(f-f^{(0)})$ to describe the system state and extract TNE and THNE information~\citep{2018-Xu-Chap2,2021-Xu-AAS,2021-Xu-AAAS,2021-Xu-JCP},
where $f$ is the distribution function and $f^{(0)}$ the equilibrium distribution function.
Therefore, the simplified Boltzmann-like equation and kinetic moment relations which are not compliant with non-equilibrium statistical physics are prohibited in the DBM.
Further, the DBM uses non-conservative kinetic moments of $(f-f^{(0)})$ to
open a phase space where the origin corresponds to the thermodynamic
equilibrium state and other points correspond to specific THNE states. The phase space and its subspaces provide a graphic geometric description of non-equilibrium states.
Evidently, each non-conservative kinetic moment of $(f-f^{(0)})$ provides its specific contribution to the overall departure from local equilibrium.
Hence, as explained above, the choice of the
non-equilibrium moments to be retained is dictated by the specific problem at hand.

Methodologically, the DBM belongs to the general framework of non-equilibrium
statistical physics, i.e.  a specific form of coarse-graining as applied to the physics
of fluids beyond the hydrodynamic level.
Specifically, the THNE effects presented by the DBM have permitted to recover
the main features of the distribution function~\citep{2014-Lin-PRE,su2022nonequilibrium},
 to capture various interfacial phenomena~\citep{2014-Lin-PRE,2016-Lai-PRE-RTI,2019-Gan-KHI}, to investigate entropy increasing mechanisms and their relative importance~\citep{zhang2016kinetic,2019-Zhang-SM-PS}, and to clarify some of the fundamental mechanisms of the fine structures of shock waves, contact discontinuities, rarefaction waves beyond the reach of molecular dynamics~\citep{2018-Gan-PRE,Qiu-POF-2020,Qiu-PRE-2021,bao2022study}.
Kinetic features, such as the unbalancing and mutual conversion of internal energy at different degrees of freedom, provide appropriate criteria for determining whether or not higher-order TNE effects should be considered in the modeling and which level of DBM should be adopted~\citep{2014-Lin-PRE,2015-Gan-SM-PS,2018-Gan-PRE}.
In plasma physics, kinetic features have been used
to distinguish plasma shock wave from shock wave in ordinary neutral flow~\citep{2022-Liu-jmse}.  These findings and observations shed light on the fundamental mechanisms of various compressible flow systems.

As for the more complicated multiphase case, in 2015, we proposed a DBM with
a more realistic equation of state (EOS) to study the HNE and TNE
manifestations induced by interparticle and gradient force during
phase transition~\citep{2015-Gan-SM-PS}.
We found that the maximum of TNE strength can be used as a physical criterion to discriminate the stages of spinodal decomposition and domain growth.
In 2019, we further found that the maximum of entropy production rate can be used as another physical criterion to discriminate the two stages~\citep{2019-Zhang-SM-PS}.
Whereas, these models are only suitable for
cases with weak non-equilibrium  effects, or cases in which the Knudsen number
is sufficiently small, including phase separation with a slow quenching rate,
shallow quenching depth, limited liquid-vapor density ratio, weak viscous
stress and heat flux, but wide interface width.

Two prominent reasons account for the insufficiency
of the model far from equilibrium.
First, constitutive relations are strictly associated with TNE measures, accounting
for the coupling of distinct lengths or time scales, and ultimately
determining the accuracy of the hydrodynamic model. Given that few kinetic
moments are satisfied by the discrete equilibrium distribution function
(DEDF), the model in the previous study~\citep{2015-Gan-SM-PS}  is only consistent with Onuki's model~\citep{Onuki-PRL-2005,Onuki-PRE-2007} for Carnahan-Starling fluid in the hydrodynamic limit with linear constitutive relations, that is,
Newton's law of viscosity and Fourier's law of heat conduction.
Nevertheless, linear constitutive relations only result from the first-order
deviation from the thermodynamic equilibrium, and are inadequate
for cases with substantial HNE and TNE effects.
Second, although higher-order TNE measures and their evolution can be extracted from this DBM~\citep{2015-Gan-SM-PS}, whether or not they are accurate enough should be carefully estimated, preserved, and improved.

In this study, we aim to develop a higher-order DBM for the multi-scale
modeling of multiphase flow systems ranging from continuum to transition
flow regimes.
Following all the inspiring works on higher-order mesoscopic models for non-equilibrium ideal-gas flows~\citep{Shan-JFM-2006,Shan-PRE-2006,Phi-PRE-2006,Shan-PRE-2019,
Shan-JFM-2021,2021-POF-Shan,Zhang-JFM-2013, Karlin-PRL-2007,Karlin-PRE-2009,Succi-PRE-2015,guo2015discrete,Sofonea-JCP-2016,
Coreixas-PRE-2017,Shu-POF-2018,Xukun-POF-2021,zhang2022discrete}
is out of the scope of the present work, although they can be extended into multiphase flow systems with the approaches listed above.
Instead,
through Chapman-Enskog multi-scale expansion, we first establish the
relationships between TNE measures and the generalized
constitutive relations for viscous stress and heat flux beyond the Burnett
level; formulate specific expressions with the second-order accuracy for
viscous stress, heat flux, and other higher-order TNE and THNE quantities
which are strictly associated with the evolution of constitutive relations.
Moreover, we determine all the fewest required kinetic moments that the DEDF
should satisfy, and construct three DBMs for multiphase flows
near-equilibrium and far-from-equilibrium by inversely solving the required
kinetic moments with efficient discrete-velocity models.
Finally, we demonstrate theoretically and numerically that the capability of
the DBM in describing TNE and THNE effects, as well as its multi-scale predictive
capability, depend on the kinetic moments specifically recovered by the DEDF.

\section{DBM for multiphase flows far-from-equilibrium}\label{sec2}

In this section, we first
reconsider what the original Boltzmann Bhatnagar-Gross-Krook (BGK) model~\citep{BGK-1954} describes
and place the BGK model frequently used for various non-equilibrium flows, in the proper theoretical
perspective for the description of non-dilute systems.
Then, we focus on the establishment of the links among Boltzmann-BGK
equation, extended hydrodynamic models and THNE phenomena;
we determine the necessary kinetic moments to measure THNE effects and
discretize phase space using a moment-matching approach.
Finally,  we present the application of DBMs to multiphase flows far from equilibrium.

\subsection{Discrete formulation of density functional kinetic theory}\label{subsec2.1}

Extended hydrodynamics or generalized hydrodynamics models consider non-equilibrium effects
 through nonlinear constitutive relations.
Correspondingly, high-order kinetic models consider non-equilibrium effects
via higher-order distribution functions deviating from equilibrium.
As a result, they are expected to be applicable to flows in the slip
and transition flow regimes where the NS equations perform poorly~\citep{Struchtrup-Book,SunQuanhua}.
A popular strategy for deriving such models is to perform a Chapman-Enskog (CE)
multi-scale expansion. In a multiphase system, the starting point is the BGK
Boltzmann equation, supplemented by an appropriate interparticle force~\citep{GLS-model-PRE2007}
\begin{equation}
\partial_{t}f+\mathbf{v} \cdot \bm{\nabla}f=-\frac{1}{\tau }[f-f^{(0)}]+I,  \label{BE}
\end{equation}
where $f^{(0)}=\frac{\rho }{2\pi T}\mathrm{exp}[-\frac{\mathbf{v}^{\ast
}\cdot \mathbf{v}^{\ast }}{2T}]$ with $\rho $, $T$, $\mathbf{v}^{\ast }$ are
the local density, temperature, and thermal velocity, respectively. Here $%
\mathbf{v}^{\ast }=\mathbf{v}-\mathbf{u}$, with $\mathbf{v}$ the particle
velocity, $\mathbf{u}$ the fluid velocity.  Finally, $I$ indicates the kinetic formulation of
interparticle force accounting for the non-ideal gas effects.

\begin{figure}
\centerline{\epsfig{file=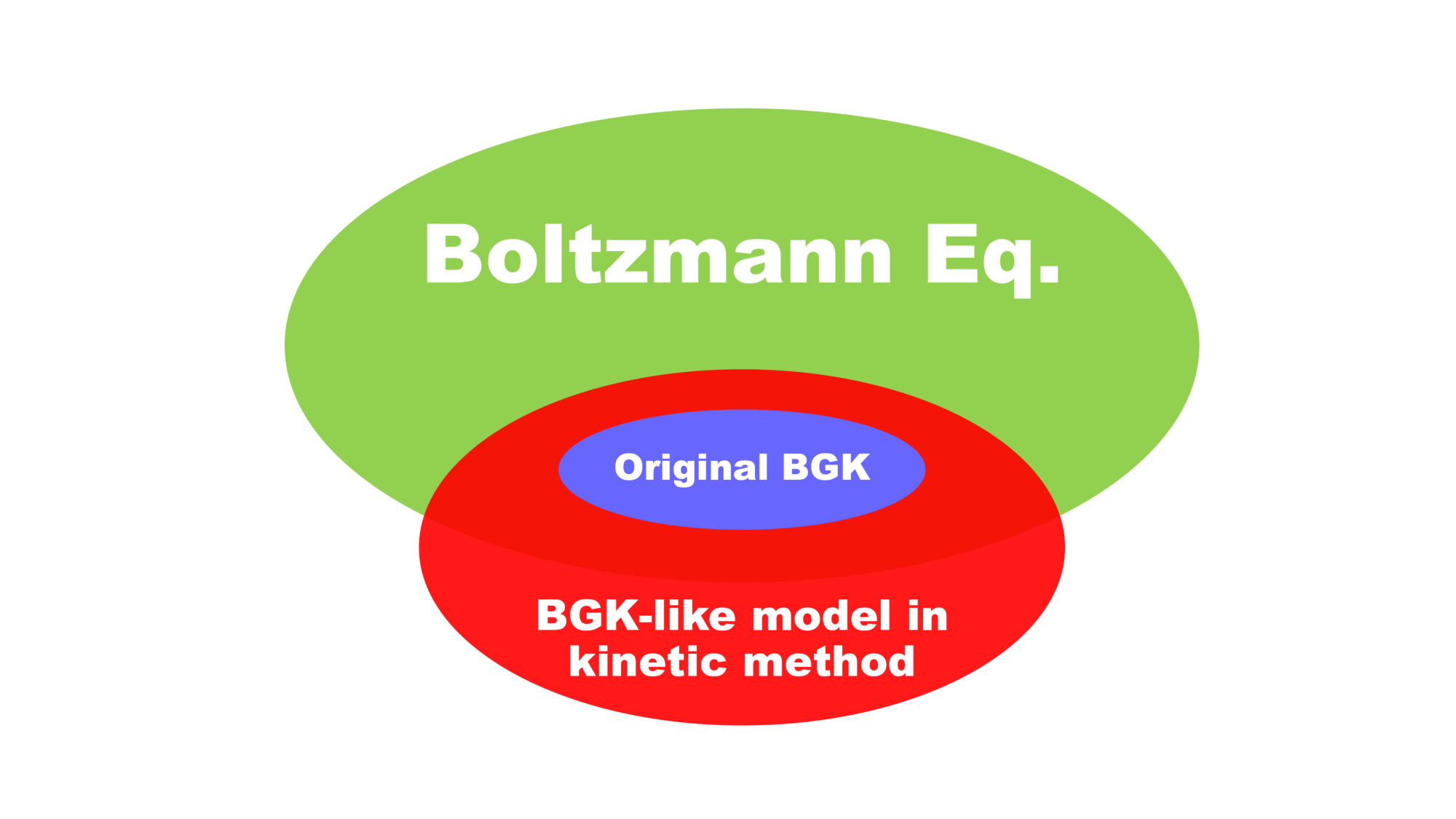,bbllx=186pt,bblly=57pt,bburx=797pt,bbury=468pt,
width=0.55\textwidth}}
\caption{Schematic diagram of the application scopes of Boltzmann equation, the original BGK model and BGK-like model in kinetic method.}
	\label{BGK}
\end{figure}

At this stage,  it is important to point out that the original BGK equation~\citep{BGK-1954}
assumes that the net effect of collisions is to relax the
velocity distribution function towards a local equilibrium distribution $f^{(0)}$,  over a
microscopic velocity-independent characteristic timescale $\tau$.
As a result,  it describes molecular collisions in an averaged statistical form, with the
only constraint of preserving mass-momentum-energy conservation, as well as
the $H$-theorem to secure convergent evolution towards a local equilibrium.
Strictly speaking, the BGK model  is valid only close to equilibrium, i.e. $Kn\ll 1$ and
$f\simeq f^{(0)}$,  thus covering only a very small portion of the full kinetic space.

To study the kinetic behavior of non-ideal fluids far from equilibrium, one needs to account for interparticle interactions beyond binary collisions, as well as non-equilibrium effects triggered by strong inhomogeneities.
This can be achieved by resorting to a mean-field theory formulation~\citep{MFT-Book}, or, more
precisely, to a density functional kinetic theory (DFKT)~\citep{DFKT}.
The role of DFKT is twofold:
(1) Include the effects of the intermolecular interaction potential, omitted by the Boltzmann equation.
(2) Extend the application scope of the BGK model to non-equilibrium phenomena beyond hydrodynamics.
In a pictorial form (see figure~\ref{BGK}), one could say that DFKT provides an ``analytic continuation" of the BGK model to regions of kinetic phase space not covered by the original Boltzmann equation itself.
Here, BGK-like refers to various families of model Boltzmann equations based on single and multi-time relaxation approximations.

In addition, the BGK-like models widely used in the study of complex fluids include
several other physical modeling improvements.
For example, relaxation times related to local macroscopic quantities~\citep{LZH-2004,LZH-2015} and collision frequency~\citep{struchtrup1997bgk}; pseudo-equilibrium distribution functions
that contain non-equilibrium information~\citep{ES-BGK,Shak-BGK,SunQuanhua,Shan-JFM-2006,2011-JCP-Meng,Watari-2016};
internal degrees of freedom~\citep{Rykov-1975} and  even quantum vibrational energy~\citep{LZH-2021-quantum}; appropriate kinetic boundary conditions~\citep{Wagner2002lees,Succi-POF-2005,toschi2005lattice,Sofonea-PRE-2005,Succi-EPL-2006,Karlin-EPL-2006,ZhangJF-PRE-2013}.
These improved models permit to capture a wider range of Knudsen numbers and a higher degree of non-equilibrium.

The multi-scale multiphase DFKT is expected to contribute to the area of materials science for energy-environment applications~\citep{Succi2020toward} and medical purposes. For instance, they may help to understand the non-equilibrium transport processes of multiphase flow in proton exchange membrane fuel cells for obtaining better performance, lower cost, emission, and noise~\citep{chen2021pore}; to investigate the non-equilibrium nucleate boiling mechanism and design suitable surface structures for enhancing nucleate boiling heat transfer efficiency~\citep{biferale2012convection,Liqing-ATE-2018,li2020enhancement,Fei-POF-2020}; to optimize heat source location~\citep{dai2020optimal} and improve the latent heat thermal energy storage rate~\citep{ren2019enhancement,tian2022artificial}; to conduct mesoscale simulations of blood flow~\citep{dupin2006multi,YWW} and generation, transmission, and diffusion of bioaerosols~\citep{WBH}.

In the present paper, as a specific example of DFKT, the mean-field force $I$ takes the following
form for a liquid-vapor system
\begin{equation}
I=-[A+\mathbf{B}\cdot \mathbf{v}^{\ast }+(C+C_{q})\mathbf{v}^{\ast }\cdot
\mathbf{v}^{\ast }]f^{(0)}\text{.}  \label{iki}
\end{equation}%
The coefficients in \eqref{iki} are designed so as to  recover the thermohydrodynamic
equations proposed by Onuki~\citep{Onuki-PRL-2005}
by using the following contributions of the force term $I$
to mass, momentum, and energy equations at the second order in Knudsen number,
 \begin{equation}
\int_{-\infty }^{\infty } Id\mathbf{v}=A+2(C+C_{q})T\text{,}
\end{equation}%
\begin{equation}
\int_{-\infty }^{\infty } I\mathbf{v}d\mathbf{v}=-\rho T\mathbf{B}\text{,}
\end{equation}%
\begin{equation}
\int_{-\infty }^{\infty } I\frac{\mathbf{v}\cdot \mathbf{v}}{2}d\mathbf{v}=-2\rho
T^{2}(C+C_{q})-\rho T\mathbf{B}\cdot \mathbf{u}\text{.}
\end{equation}
As a result, $A=-2(C+C_{q})T$, indicates that the forcing term guarantees mass conservation;
$\mathbf{B}=\frac{1}{\rho T}[\bm{\nabla}(P^{\text{CS}}-\rho T)+\bm{\nabla}\cdot \bm{\Lambda}]$,
describes two contributions of $I$ to the fluid momentum.
One comes from the difference between non-ideal gas EOS and ideal gas EOS, and the other
from contribution of density gradient to pressure tensor $\bm{\Lambda}=K\bm{\nabla}\rho \bm{\nabla}\rho -K(\rho \nabla ^{2}\rho +\left\vert \bm{\nabla}\rho \right\vert ^{2}/2)\mathbf{I}-[\rho T\bm{\nabla}\rho \cdot\bm{\nabla}(K/T)]\mathbf{I}$,
with $\mathbf{I}$ the unit tensor, $K$ the surface tension
coefficient. Here we choose the Carnahan-Starling EOS~\citep{carnahan1969equation} as the non-ideal one
$P^{\text{CS}}=\rho T\frac{1+\eta +\eta ^{2}-\eta ^{3}}{(1-\eta
)^{3}}-a\rho ^{2}$ with $\eta =b\rho /4$,
which offers a more accurate representation
for hard sphere interactions than the van der
Waals EOS. More realistic EOS, such as the
Redlich-Kwong~\citep{redlich1949thermodynamics}, the Peng-Robinson EOS~\citep{peng1976new} could be
conveniently incorporated into the DBM through modifying the extra force
term $I$ for modeling improvement.
$C=\frac{1}{2\rho T^{2}}%
\{(P^{CS}-\rho T)\bm{\nabla}\cdot \mathbf{u}+{\bm{\Lambda}}\colon \bm{\nabla}%
\mathbf{u}+a\rho ^{2}\bm{\nabla}\cdot \mathbf{u}+K[-\frac{1}{2}(\bm{\nabla}%
\rho \cdot \bm{\nabla}\rho )\bm{\nabla}\cdot \mathbf{u}-\rho \bm{\nabla}\rho
\cdot \bm{\nabla}(\bm{\nabla}\cdot \mathbf{u})-\bm{\nabla}\rho \cdot %
\bm{\nabla}\mathbf{u}\cdot \bm{\nabla}\rho ]\}$, represents a partial contribution of
$I$ to energy.
The term $C_{q}=\frac{1}{2\rho T^{2}}
\bm{\nabla}\cdot \lbrack 2q\rho T\bm{\nabla}T]$, permits to tune the
heat conductivity independently of the dynamic viscosity (non-unit Prandtl number).

Constructing the density, momentum and energy kinetic moments of
 (\ref{BE}) and recomposing the time derivative, result in the following
extended hydrodynamics equations:
\begin{equation}
\partial _{t}\rho +\bm{\nabla}\cdot (\rho \mathbf{u})=0\text{,}  \label{NS1}
\end{equation}%
\begin{equation}
\partial _{t}(\rho \mathbf{u})+\bm{\nabla}\cdot (\rho \mathbf{uu}+\bm{
\Pi }+\bm{\Delta}_{2})=0\text{,}  \label{NS2}
\end{equation}%
\begin{equation}
\partial _{t}e_{T}+\bm{\nabla}\cdot \lbrack e_{T} \mathbf{u}+\bm{\Pi}\cdot
\mathbf{u}+\bm{\Delta}_{3,1}+\kappa_{2}\bm{\nabla}T]=0\text{,}  \label{NS3}
\end{equation}%
where $\bm{\Pi}=P^{\text{CS}}\mathbf{I}+\bm{\Lambda }$ is the
nonviscous stress, $e_{T}=\rho T-a\rho ^{2}+K|\bm{\nabla}\rho |^{2}/2+\rho
u^{2}/2$ is the total energy density, $\kappa _{2}=2\rho Tq$ is an
additional heat conductivity originated from $C_{q}$ and designed to tune
the Prandtl number. $\bm{\Delta}_{2}$ and $\bm{\Delta}_{3,1}$ are typical
non-equilibrium  measures related to the generalized constitutive relations.

Without loss of generality, two sets of non-equilibrium  measures can be
defined~\citep{2012-Xu-FOP,2015-Gan-SM-PS},
\begin{equation}
\bm{\Delta}_{m,n}=\mathbf{M}_{m,n}(f-f^{(0)})=\int_{-\infty }^{\infty } (\frac{1}{2})^{1-\delta
_{m,n}}(f^{(1)}+f^{(2)}+...){\underbrace{{\mathbf{vv}}\cdots {\mathbf{v}}}%
_{n}(\mathbf{v\cdot v})}^{\frac{m-n}{2}}d\mathbf{v}\text{,}
\label{THNE}
\end{equation}%
\begin{equation}
\bm{\Delta}_{m,n}^{\ast }=\mathbf{M}_{m,n}^{\ast }(f-f^{(0)})=\int_{-\infty }^{\infty } (\frac{1}{%
2})^{1-\delta _{m,n}}(f^{(1)}+f^{(2)}+...){\underbrace{{\mathbf{v}}^{\ast }{%
\mathbf{v}}^{\ast }\cdots {\mathbf{v}}^{\ast }}_{n}(\mathbf{v}}^{\ast }{%
\mathbf{\cdot v}}^{\ast }{)}^{\frac{m-n}{2}}d\mathbf{v}\text{,}
\label{TNE}
\end{equation}%
where $\bm{\Delta }_{m,n}$ and $\bm{\Delta }_{m,n}^{\ast }$ are the $m$-th
order tensors contracted to $n$-th order ones, and $\delta_{m,n}$ is the
Kronecker delta function.
Here $(m-n)/2$ is an integer, when $m=n$, $\bm{\Delta }_{m,n}$ and $\bm{\Delta }%
_{m,n}^{\ast }$ are simplistically referred to as $\bm{\Delta }_{m}$ and $%
\bm{\Delta }_{m}^{\ast }$, respectively. Here $\bm{\Delta}_{m,n}$ describes
the combination effects of HNE and TNE, which are usually called the
thermo-hydrodynamic non-equilibrium  (THNE) effects, $\bm{\Delta }_{m,n}^{\ast
}$ reflects molecular individualism on top of organized collective motion,
describing only the TNE effects; $\bm{\Delta}_{m,n}-\bm{\Delta}_{m,n}^{\ast }$ depicts the HNE effects, as a supplemental description of
macroscopic quantity gradients. $\mathbf{M}_{m,n}$ ($\mathbf{M}_{m,n}^{\ast
} $) is the non-central (central) kinetic moment, $f^{(j)}$ represents the $%
j $-th order derivation from $f^{(0)}$.
The decomposition relations among THNE and TNE measures are derived and displayed in Appendix \ref{appendix1},
\begin{equation}
\begin{array}{c}
\bm{\Delta}_{2}=\bm{\Delta}_{2}^{\ast }\text{, } \\
\bm{\Delta}_{3,1}{\mathbf{=}}\bm{\Delta}_{3,1}^{\ast }+\bm{\Delta}_{2}^{\ast
}\cdot {\mathbf{u}}\text{,} \\
\bm{\Delta}_{3}=\bm{\Delta}_{3}^{\ast }+(u_{\alpha }\Delta _{2\beta \gamma
}^{\ast }+u_{\beta }\Delta _{2\alpha \gamma }^{\ast }+u_{\gamma }\Delta
_{2\alpha \beta }^{\ast })\mathbf{e}_{\alpha }\mathbf{e}_{\beta }\mathbf{e}%
_{\gamma }\text{,} \\
\bm{\Delta}_{4,2}=\bm{\Delta}_{4,2}^{\ast }+\bm{\Delta}_{3,1}^{\ast }{%
\mathbf{u+u}}\bm{\Delta}_{3,1}^{\ast }+\bm{\Delta}_{3}^{\ast }\cdot {\mathbf{%
u}}+\bm{\Delta}_{2}^{\ast }\cdot {\mathbf{uu}}+\mathbf{u}\bm{\Delta}%
_{2}^{\ast }\cdot {\mathbf{u+}}\frac{u^{2}}{2}\bm{\Delta}_{2}^{\ast }\text{.}
\end{array}
\label{THNE-TNE}
\end{equation}
Here $\bm{\Delta }_{2}^{\ast}$ and $\bm{\Delta }_{3,1}^{\ast }$ are also referred to as the
non-organized moment flux (NOMF) and non-organized energy flux (NOEF),
respectively.
Compared with the NS and Burnett equations, $\bm{\Delta}%
_{2}^{\ast }$ and $\bm{\Delta }_{3,1}^{\ast }$ correspond to more
generalized viscous stress and heat flux, which contain all hierarchical TNE
effects induced by $(f-f^{(0)})$, instead of only $f^{(1)}$ at the NS level
and $f^{(1)}+f^{(2)}$ at the Burnett level.
The links among BE, the extended hydrodynamic equations, and TNE measures
have been established so far. $\bm{\Delta }_{3}^{\ast}$ and $\bm{\Delta }_{4,2}^{\ast}$ are flux of viscous stress and flux of heat flux, respectively.

However, $\bm{\Delta}_{2}^{\ast }$ and $\bm{\Delta}_{3,1}^{\ast }$ are
primarily unknown. To obtain the explicit constitutive relations, we perform
Chapman-Enskog expansion on both sides of \eqref{BE} by introducing
expansions
\begin{equation}
f=f^{(0)}+f^{(1)}+f^{(2)}+\cdots \text{,}  \label{CE-1}
\end{equation}%
\begin{equation}
\partial _{t}=\partial _{t_{1}}+\partial _{t_{2}}+\cdots \text{,}
\label{CE-2}
\end{equation}%
\begin{equation}
\bm{\nabla}=\bm{\nabla}_{1}\text{, \ }I=I_{1}\text{,}  \label{CE-3}
\end{equation}%
where $\partial _{t_{j}}$, $\bm{\nabla}_{j}$, and $I_{j}$ are $j$-th order
terms in Knudsen number $\epsilon $. Substituting equations \eqref{CE-1}-\eqref{CE-3} into \eqref{BE} and equating terms that have the same orders in
$\epsilon $ give
\begin{equation}
\epsilon ^{1}:{\partial }_{t_{1}}{f^{(0)}}+\bm{\nabla}_{1}\cdot (f^{(0)}%
\mathbf{v})=-\frac{1}{\tau }f^{{(1)}}+I_{1}\text{,}  \label{1st}
\end{equation}%
\begin{equation}
\epsilon ^{2}:{\partial }_{t_{2}}{f^{(0)}}+{\partial }_{t_{1}}{f^{(1)}}+%
\bm{\nabla}_{1}\cdot (f^{(1)}\mathbf{v})=-\frac{1}{\tau }f^{{(2)}}\text{.}
\label{2nd}
\end{equation}%
Performing the velocity moments with the collision invariant vector ($1$, $%
\mathbf{v}$, $\frac{v^{2}}{2}$) on the two sides of \eqref{1st}, leads
to the useful relations between the temporal derivative $\partial _{t_{1}}$
and the spatial derivative $\bm{\nabla}_{1}$ at $\epsilon ^{1}$ level
\begin{equation}
{\partial }_{t_{1}}\rho =-\bm{\nabla}_{1}\cdot (\rho \mathbf{u})\text{,}
\label{rho_t1}
\end{equation}%
\begin{equation}
{\partial }_{t_{1}}\mathbf{u}=-T\mathbf{B}-\bm{\nabla}_{1}T-\frac{T}{\rho }%
\bm{\nabla}_{1}\rho -\mathbf{u}\cdot \bm{\nabla}_{1}\mathbf{u}\text{,}
\label{u_t1}
\end{equation}%
\begin{equation}
{\partial }_{t_{1}}T=-2(C+C_{q})T^{2}-T\bm{\nabla}_{1}\cdot \mathbf{u-u}%
\cdot \bm{\nabla}_{1}T\text{.}  \label{T_t1}
\end{equation}%

From \eqref{1st}, we can
obtain $f^{{(1)}}=-\tau \lbrack {\partial }_{t_{1}}f{^{(0)}}+\bm{\nabla}%
_{1}\cdot (f^{(0)}\mathbf{v})-I_{1}]$.
Using \eqref{rho_t1}-\eqref{T_t1} and after some straightforward
but rather tedious algebraic manipulation, we acquire the following
relations between thermodynamic forces and fluxes
\begin{equation}
\bm{\Delta }_{2}^{\ast (1)}=\int_{-\infty }^{\infty } f^{(1)}\mathbf{v}^{\ast }\mathbf{v}^{\ast }d%
\mathbf{v}=-\mu \lbrack \bm{\nabla}\mathbf{u}+(\bm{\nabla}\mathbf{u})^{T}-%
\mathbf{I}\bm{\nabla}\cdot \mathbf{u}]\text{,}  \label{D2}
\end{equation}%
\begin{equation}
\bm{\Delta }_{3,1}^{\ast (1)}=\int_{-\infty }^{\infty } \frac{1}{2}f^{(1)}\mathbf{v}^{\ast }\cdot
\mathbf{v}^{\ast }\mathbf{v}^{\ast }d\mathbf{v}=-\kappa _{1}\bm{\nabla}T%
\text{,}  \label{D31}
\end{equation}%
Equations \eqref{D2}-\eqref{D31} indicate that the first-orders of NOMF and
NOEF are exactly the viscous stress tensor and the heat flux at the NS
level, respectively; $\mu =\rho T\tau $ is the viscosity coefficient and $%
\kappa _{1}=$ $2\rho T\tau $ the heat conductivity.
Similarly, we have
\begin{equation}
\bm{\Delta}_{3}^{\ast (1)}=\int_{-\infty }^{\infty }f^{(1)}{\mathbf{v}}%
^{\ast }{\mathbf{v}}^{\ast }{\mathbf{v}}^{\ast }d\mathbf{v}=-\rho T\tau
(\partial _{\alpha }T\delta _{\beta \gamma }+\partial _{\beta }T\delta
_{\alpha \gamma }+\partial _{\gamma }T\delta _{\alpha \beta })\mathbf{e}%
_{\alpha }\mathbf{e}_{\beta }\mathbf{e}_{\gamma }\text{.}
\end{equation}
\begin{equation}
\bm{\Delta}_{4,2}^{\ast (1)}=\int_{-\infty }^{\infty }f^{(1)}{\mathbf{v}}%
^{\ast }{\mathbf{v}}^{\ast }\frac{{\mathbf{v}}^{\ast }\cdot {\mathbf{v}}%
^{\ast }}{2}d\mathbf{v}=-3\rho T^{2}\tau \lbrack \bm{\nabla}\mathbf{u}+(%
\bm{\nabla}\mathbf{u})^{T}-\mathbf{I}\bm{\nabla}\cdot \mathbf{u}].
\end{equation}

Deeper TNE effects can be accessed by considering the contributions of $
f^{(2)}$, and clarifying relations between $\partial _{t_{2}}$ and $\bm{\nabla}_{1}$ at $\epsilon ^{2}$ level according to similar procedures listed in \eqref{rho_t1}-\eqref{T_t1}.
Specifically, from \eqref{2nd}, we acquire
\begin{eqnarray}
f^{{(2)}} &=&-\tau \lbrack {\partial }_{t_{2}}f{^{(0)}}+{\partial }_{t_{1}}f{%
^{(1)}}+\bm{\nabla}_{1}\cdot (f^{(1)}\mathbf{v})] \\
&=&-\tau {\partial }_{t_{2}}f{^{(0)}}+\tau ^{2}{\partial }_{t_{1}}^{2}f{%
^{(0)}}+\tau ^{2}{\partial }_{t_{1}}[\bm{\nabla}_{1}\cdot (f^{(0)}\mathbf{v}%
)]-\tau ^{2}{\partial }_{t_{1}}I_{1}  \notag \\
&&+\tau ^{2}\bm{\nabla}_{1}\cdot \lbrack {\partial }_{t_{1}}f{^{(0)}}\mathbf{%
v}+\bm{\nabla}_{1}\cdot (f^{(0)}\mathbf{vv})-I_{1}\mathbf{v}].  \notag
\end{eqnarray}%
It is clear that $f^{(2)}$ can also be expressed in terms of $f^{(0)}$.
As a result, non-conservative non-equilibrium moments, can finally
resort to those of $f^{(0)}$, which triggers the requirement of the more
higher-order kinetic moments of $f^{(0)}$.
The second order non-organized moment flux reads
\begin{equation}
\bm{\Delta}_{2}^{\ast (2)}=\mathbf{M}_{2}^{\ast }(f^{(2)})=\int_{-\infty
}^{\infty }f^{(2)}{\mathbf{v}}^{\ast }{\mathbf{v}}^{\ast }d\mathbf{v}.
\end{equation}%
To obtain an explicit expression, relations between $\partial _{t_{2}}$ and $%
\bm{\nabla}_{1}$ at $\epsilon ^{2}$ level should be clarified. Performing
the velocity moments with the collision invariant vector ($1$, $\mathbf{v}$,
$\frac{v^{2}}{2}$) on the two sides of equation (\ref{2nd}), leads
to the useful relations between the temporal derivative $\partial _{t_{2}}$
and the spatial derivative $\bm{\nabla}_{1}$ at $\epsilon ^{2}$ level
\begin{equation}
{\partial }_{t_{2}}\rho =0\text{,}  \label{rho_t2}
\end{equation}%
\begin{equation}
\rho {\partial }_{t_{2}}\mathbf{u}=-\bm{\nabla}_{1}\cdot \bm{\Delta}%
_{2}^{(1)}\text{,}  \label{u_t2}
\end{equation}%
\begin{equation}
{\partial }_{t_{2}}(\rho T+\frac{1}{2}\rho u^{2})=-\bm{\nabla}_{1}\cdot %
\bm{\Delta}_{3,1}^{(1)}\text{.}  \label{T_t2}
\end{equation}%
Using Eqs. (\ref{rho_t2})-(\ref{T_t2}) to replace the temporal derivative $%
\partial _{t_{2}}$ by spatial derivative $\bm{\nabla}_{1}$ at $\epsilon ^{2}$
level, and after some laborious algebraic manipulation, we acquire expression
for $\bm{\Delta}_{2}^{\ast (2)}$. Likewise, analytical
expressions for $\bm{\Delta}_{3,1}^{\ast (2)}$, $\bm{\Delta}_{3}^{\ast (2)}$
and $\bm{\Delta}_{4,2}^{\ast (2)}$ can be obtained, as illustrated in Appendix \ref{appendix2}. The derivation of these expressions is a tough task, so we resort to
software, Maple 18.


Furthermore, multiplying \eqref{BE} by $\mathbf{v}^{\ast }\mathbf{v}%
^{\ast }$ and $\frac{1}{2}\mathbf{v}^{\ast }\cdot \mathbf{v}^{\ast }\mathbf{v%
}^{\ast }$, and integrating over the whole velocity space, we obtain the
time evolution equations for viscous stress and heat flux,
\begin{equation}
{\partial }_{t}\bm{\Delta }_{2}^{\ast }+{\partial }_{t}\mathbf{M}_{2}^{\ast
}(f^{(0)})+\bm{\nabla}\cdot \lbrack \mathbf{M}_{3}^{\ast }(f^{(0)})+\mathbf{M%
}_{2}^{\ast }(f^{(0)})\mathbf{u}+\bm{\Delta }_{3}^{\ast }+\bm{\Delta }%
_{2}^{\ast }\mathbf{u}]=-\frac{1}{\tau }\bm{\Delta }_{2}^{\ast }+\mathbf{M}%
_{2}^{\ast }(I)\text{,}  \label{E1}
\end{equation}%
\begin{equation}
{\partial }_{t}\bm{\Delta }_{3,1}^{\ast }+{\partial }_{t}\mathbf{M}%
_{3,1}^{\ast }(f^{(0)})+\bm{\nabla}\cdot \lbrack \mathbf{M}_{4,2}^{\ast
}(f^{(0)})+\mathbf{M}_{3,1}^{\ast }(f^{(0)})\mathbf{u}+\bm{\Delta }%
_{4,2}^{\ast }+\bm{\Delta }_{3,1}^{\ast }\mathbf{u}]=-\frac{1}{\tau }%
\bm{\Delta }_{3,1}^{\ast }+\mathbf{M}_{3,1}^{\ast }(I)\text{.}  \label{E2}
\end{equation}
Equations \eqref{E1} and \eqref{E2} suggest that obtaining the expressions
of higher-order TNE quantities, such as $\bm{\Delta}_{3}^{\ast }$ and $%
\bm{\Delta }_{4,2}^{\ast }$,
enhances our understanding of the evolution mechanisms of nonlinear
constitutive relations.


To summarize, higher-order constitutive relations of second-order accuracy for viscous stress and heat transfer can be formulated as $\bm{\Delta }_{2}^{\ast }=\bm{\Delta }_{2}^{\ast (1)}+\bm{\Delta }%
_{2}^{\ast (2)}$, $\bm{\Delta }_{3,1}^{\ast }=\bm{\Delta }_{3,1}^{\ast (1)}+%
\bm{\Delta }_{3,1}^{\ast (2)}$, which are expected to noticeably improve the
macroscopic modeling.
Meanwhile, higher-order TNE measures with second-order accuracy, $\bm{\Delta }_{3}^{\ast }=\bm{\Delta }_{3}^{\ast (1)}+\bm{\Delta }_{3}^{\ast (2)}$, $%
\bm{\Delta }_{4,2}^{\ast }=\bm{\Delta }_{4,2}^{\ast (1)}+\bm{\Delta }%
_{4,2}^{\ast (2)}$,
offer more abundant non-equilibrium scenarios from different aspects and are useful in elucidating the evolutions of higher-order constitutive relations.

In fact, the theoretical expressions of $\bm{\Delta }_{3}^{\ast}$ and $\bm{\Delta }_{4,2}^{\ast}$ are also constitutive relations, which are omitted in the NS equations.
In the case of quasi-continuous and near-equilibrium, it is not a big problem if they are omitted. However, as the degree of discretization and thermodynamic non-equilibrium increase, ignoring $\bm{\Delta }_{3}^{\ast}$ and $\bm{\Delta }_{4,2}^{\ast}$ leads to substantial errors.
Furthermore, the analytical expressions for
THNE with second-order  accuracy, $\bm{\Delta}_{2}=\bm{\Delta}_{2}^{(1)}+\bm{\Delta}_{2}^{(2)}$, $\bm{\Delta}_{3,1}=\bm{\Delta}_{3,1}^{(1)}+\bm{\Delta}_{3,1}^{(2)}$, $\bm{\Delta}_{3}=\bm{\Delta}_{3,1}^{(1)}+\bm{\Delta}_{3,1}^{(2)}$, and $\bm{\Delta}_{4,2}=\bm{\Delta}_{4,2}^{(1)}+\bm{\Delta}_{4,2}^{(2)}$ can be acquired
from \eqref{THNE-TNE}.

\subsection{Discretization of particle velocity space}\label{subsec2.2}

The velocity space discretization is a critical step in DBM modeling. The
above expressions for TNE measures (see Appendix \ref{appendix2}) have been derived from
the \eqref{BE} with a continuous equilibrium distribution function $%
f^{(0)} $. To conduct numerical simulation, it is required to construct a
discrete-velocity model (DVM) $\mathbf{v}_{i}$ to discretize the particle
velocity space, and formulate the DEDF $f_{i}^{(0)}$ satisfying all the
needed kinetic moments, where $i=1,2,3,...,N$, with $N$ the total number of
discrete velocities. Chapman-Enskog analysis demonstrates that $f_{i}^{(0)}$
should satisfy the following seven kinetic moments~\citep{2015-Gan-SM-PS}: $%
\mathbf{M}_{0}$, $\mathbf{M}_{1}$, $\mathbf{M}_{2,0}$, $\mathbf{M}_{2}$, $%
\mathbf{M}_{3,1}$, $\mathbf{M}_{3}$, $\mathbf{M}_{4,2}$ in the recovery of
the targeted hydrodynamic equations at the NS level.
In cases with fixed specific heat-ratio,
$\mathbf{M}_{0}$, $\mathbf{M}_{1}$, $\mathbf{M}_{2}$, $\mathbf{M}_{3}$, $%
\mathbf{M}_{4,2}$ are independent. In the recovery of the targeted
hydrodynamic equations at the Burnett level, $f_{i}^{(0)}$ should further
satisfy $\mathbf{M}_{4}$ and $\mathbf{M}_{5,3}$~\citep{2018-Gan-PRE}. In the
recovery of the targeted hydrodynamic equations beyond the Burnett level and
description of all the TNE measures displayed in Appendix \ref{appendix2} with
second-order accuracy, $f_{i}^{(0)}$ should satisfy more higher-order
non-equilibrium  moments, as listed in Table \ref{table1}. Among them, $\mathbf{M}_{0}$, $%
\mathbf{M}_{1}$, $\mathbf{M}_{2}$, $\mathbf{M}_{3}$, $\mathbf{M}_{4}$, $%
\mathbf{M}_{5}$, $\mathbf{M}_{6,4}$, and $\mathbf{M}_{7,3}$ are independent
(see Appendix \ref{appendix3}).

\begin{table}
\begin{center}
\begin{tabular}[t]{lcccc}
TNE & Required & higher-order & kinetic & moments \\ \hline
$\bm{\Delta}_{3}^{*(1)}$ & $\textcolor{blue}{\mathbf{M}_{4}}$ & $
\textcolor{blue}{\mathbf{M}_{5}}$ &  &  \\
$\bm{\Delta}_{3,1}^{*(1)}$ & $\mathbf{M}_{5,1}$ &  &  &  \\
$\bm{\Delta}_{4,2}^{(1)}$ & $\mathbf{M}_{5,3}$ & $\mathbf{M}_{6,2}$ &  &  \\ 
$\bm{\Delta}_{2}^{*(2)}$ & $\mathbf{M}_{4}$ & $\mathbf{M}_{5,3}$ & &  \\
$\bm{\Delta}_{3}^{*(2)}$ & $\mathbf{M}_{4}$ & $\mathbf{M}_{5}$ & $%
\mathbf{M}_{5,3}$ & $\textcolor{blue}{\mathbf{M}_{6,4}}$ \\
$\bm{\Delta}_{3,1}^{*(2)}$ & $\mathbf{M}_{5,1}$ & $\mathbf{M}_{5,3}$
& $\mathbf{M}_{6,2}$ &  \\
$\bm{\Delta}_{4,2}^{*(2)}$ & $\mathbf{M}_{5,3}$ & $\mathbf{M}_{6,2}$
& $\mathbf{M}_{6,4}$ & $\textcolor{blue}{\mathbf{M}_{7,3}}$ \\ \hline
\end{tabular}
\caption{Higher-order kinetic moment relations required for deriving the
analytical expressions of TNE measures, where the blue ones are
independent constraint relations.}
\label{table1}
\end{center}
\end{table}

\begin{figure}
{\centerline{\epsfig{file=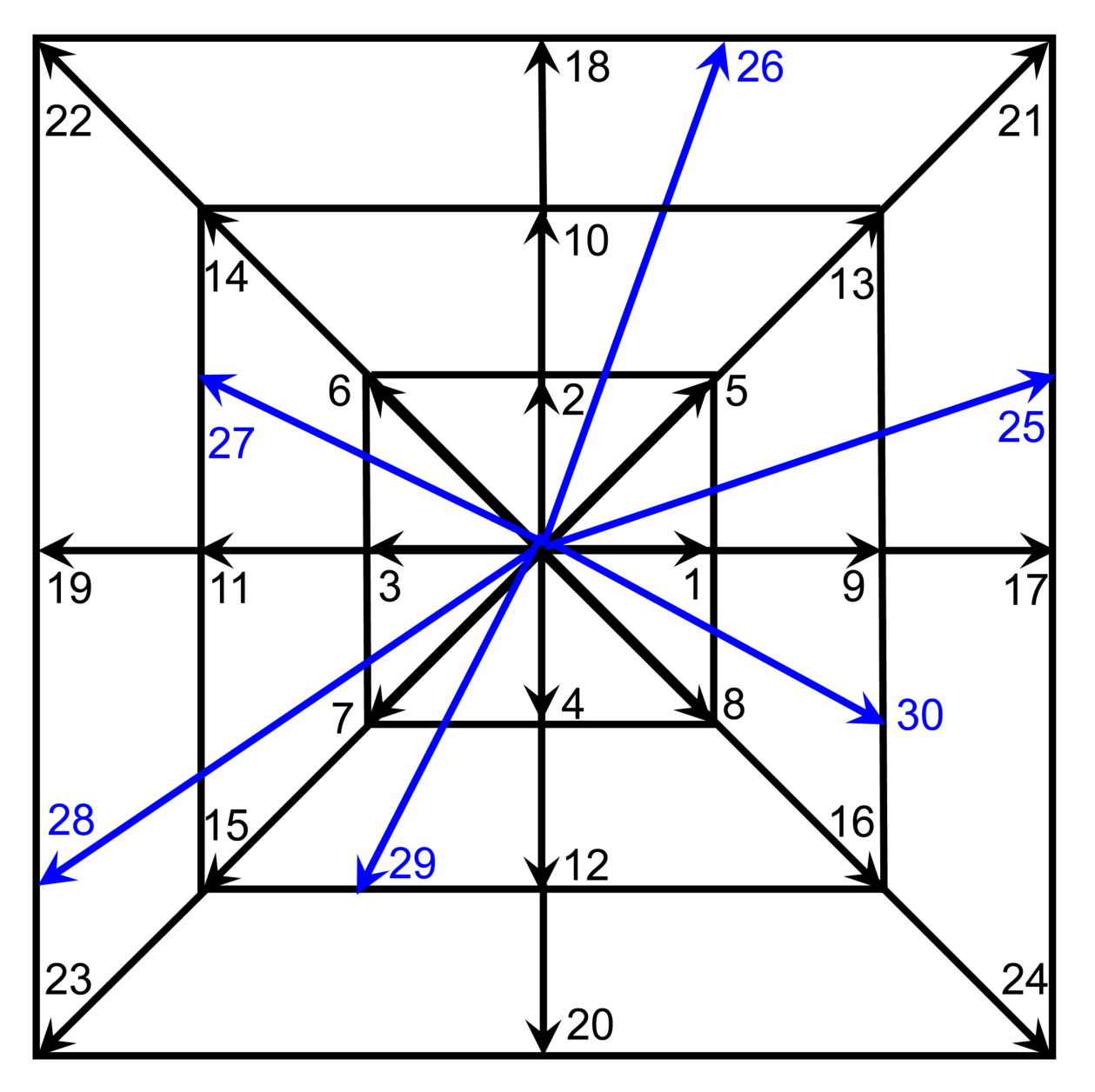,bbllx=13pt,bblly=10pt,bburx=590pt,bbury=591pt,
				width=0.45\textwidth,clip=}}}
\caption{Schematic of the D2V30 discrete-velocity model,
where $\mathbf{v} _{1},\mathbf{v}_{2},...,\mathbf{v}_{24}$ are symmetric
vectors, and $\mathbf{\ v}_{25},\mathbf{v}_{26},...,\mathbf{v}_{30}$ are non-symmetric ones, designed to guarantee the existence of $\bm{\Psi}{^{-1}}$. }
\label{DVM}
\end{figure}

Here, \textquotedblleft satisfy\textquotedblright\ means that the
required kinetic moments of $f_{i}^{(0)}$, originally in integral form, can
be calculated in discrete summation form
\begin{equation}
\bm{\Phi}_{n}(f^{(0)})=\int_{-\infty }^{\infty } f^{(0)}\bm{\Psi }d\mathbf{v=}%
\sum\limits_{i}f_{i}^{(0)}\bm{\Psi }_{i}\text{,}  \label{Moments}
\end{equation}%
with $\bm{\Psi }=[1,\mathbf{v},\mathbf{vv},\mathbf{vvv},\mathbf{vvvv},%
\mathbf{vvvvv},\frac{1}{2}\mathbf{v\cdot vvvvv},\frac{1}{2}\mathbf{(v\cdot v)%
}^{2}\mathbf{vvv}]^{T}$. Equation \eqref{Moments} can be rewritten in a
matrix form
\begin{equation}
\bm{\Phi }_{n}=\bm{\Psi }\cdot \mathbf{f}^{(0)}\text{,}  \label{Feq_1}
\end{equation}%
where $\bm{\Phi}_{n}=(\mathbf{M}_{0},\mathbf{M}_{1},\mathbf{M}_{2},
\mathbf{M}_{3},\mathbf{M}_{4},\mathbf{M}_{5},\mathbf{M}_{6,4},\mathbf{M}%
_{7,3})^{T}=(M_{0},M_{1x},M_{1y},...,M_{7,3yyy})^{T} $ is the set of moments
of $f^{(0)}$. $\mathbf{f}%
^{(0)}=(f_{1}^{(0)},f_{2}^{(0)},...,f_{30}^{(0)})^{T}$, $\bm{\Psi}=(\mathbf{C}_{1},\mathbf{C}_{2},\cdots,\mathbf{C}_{30})$ is a $30\times 30$
matrix bridging the DEDF and the kinetic moments with $\mathbf{C}%
_{i}=(1,v_{ix},v_{iy},...,\frac{1}{2}v^{4}v_{ix}^{3},\frac{1}{2}%
v^{4}v_{ix}^{2}v_{iy},\frac{1}{2}v^{4}v_{ix}v_{iy}^{2},\frac{1}{2}%
v^{4}v_{iy}^{3})^{T}$. Naturally, $\mathbf{f}^{(0)}$ can be calculated as
follows~\citep{2013-Gan-EPL},
\begin{equation}
\mathbf{f}^{(0)}=\bm{\Psi}{^{-1}} \cdot \bm{\Phi }_{n}\text{,}
\label{f0}
\end{equation}%
where $\bm{\Psi}{^{-1}}$ is the inverse matrix of $\bm{\Psi}$. Here, a
two-dimensional DVM with $30$ discrete velocities, schematically drawn in
figure \ref{DVM}, is used to discretize the velocity space and to ensure the existence
of $\bm{\Psi}{^{-1}}$
\begin{equation}
(v_{ix},v_{iy})=\left\{
\begin{array}{cc}
\text{cyc}:c(\pm 1,0) & \text{for}\quad 1\leq i\leq 4 \\
c(\pm 1,\pm 1) & \text{for}\quad 5\leq i\leq 8 \\
\text{cyc}:2c(\pm 1,0) & \text{ \ for}\quad 9\leq i\leq 12 \\
2c(\pm 1,\pm 1) & \text{ \ \ for}\quad 13\leq i\leq 16 \\
\text{cyc}:3c(\pm 1,0) & \text{ \ \ for}\quad 17\leq i\leq 20 \\
3c(\pm 1,\pm 1) & \text{ \ \ for}\quad 21\leq i\leq 24 \\
c(3,1),c(1,3) & \text{for}\quad i=25,26 \\
c(-2,1),c(-3,-2) & \text{for}\quad i=27,28 \\
c(-1,-2),c(2,-1) & \text{for}\quad i=29,30
\end{array}%
\right. \text{,}
\end{equation}%
where \textquotedblleft cyc" indicates cyclic permutation. The selection of
$\mathbf{v}_{25},...,\mathbf{v}_{30}$ is highly flexible and thus guarantees
the existence of $\bm{\Psi}{^{-1}}$ and optimizes the stability of the model.
Then the discrete coarse-grained evolution equation reads
\begin{equation}
\partial_{t}f_i+\mathbf{v}_i \cdot \bm{\nabla}f_i=-\frac{1}{\tau }[f_i-f^{(0)}_i]+I_i.  \label{BE2}
\end{equation}

\begin{table}
\begin{center}
\begin{tabular}[t]{lcccc}
DVMs & & & Satisfied  independent  kinetic  moments \\ 
D2V13 & & & $\mathbf{M}_{0}$, $\mathbf{M}_{1}$, $\mathbf{M}_{2}$, $\mathbf{M}_{3}$, $\mathbf{M}_{4,2}$ \\ 
D2V15 & & & $\mathbf{M}_{0}$, $\mathbf{M}_{1}$, $\mathbf{M}_{2}$, $\mathbf{M}_{3}$, $\mathbf{M}_{4,2}$, $\mathbf{M}_{5,1}$ \\ 
D2V16  & & & $\mathbf{M}_{0}$, $\mathbf{M}_{1}$, $\mathbf{M}_{2,0}$, $\mathbf{M}_{2}$, $\mathbf{M}_{3,1}$, $\mathbf{M}_{3}$, $\mathbf{M}_{4,2}$ \\ 
D2V30 & & & $\mathbf{M}_{0}$, $\mathbf{M}_{1}$, $\mathbf{M}_{2}$, $\mathbf{M}_{3}$, $\mathbf{M}_{4}$, $\mathbf{M}_{5}$, $\mathbf{M}_{6,4}$, $\mathbf{M}_{7,3}$ \\ 
\end{tabular}%
\caption{Discrete-velocity models and corresponding independent kinetic moments.}
\label{table2}
\end{center}
\end{table}

It is necessary to point out that:

(i) The most important step in DBM
modeling and simulation is the calculation of $f_{i}^{(0)}$. The Maxwell
distribution satisfies an infinite number of moment relations. Requiring $%
f_{i}^{(0)}$ to satisfy all the moment relations is impractical. Which
moment relations are necessary
depends on the kinetic properties to be studied, a choice that is typically dictated by the intensity of the non-equilibrium effects to be described.

(ii) The capability of describing non-equilibrium effects relies on the
kinetic moment relations that the DBM correctly retrieved. Therefore, by
satisfying different kinetic moments, we can systematically construct DBMs
for TNE effects at various levels that correspond to different
coarse-grained models and own different application ranges. For instance,
besides the D2V30 DBM, we propose D2V13, D2V15, and D2V16 multiphase models,
and examine their performances in the subsequent section. Here, the D2V13
(D2V15) model adopts the former 13 (15) discrete velocities of the D2V30
model. D2V16 model indicates the one proposed in~\citep{2013-Gan-EPL}
for compressible flows with a flexible specific heat ratio. While considering
only the translational degrees of freedom, the D2V16 model will degenerate
into the D2V13 model. Independent kinetic moments retrieved by these models are
exhibited in Table \ref{table2}.

(iii) For the same set of grids, the computational cost of the model is proportional to the number of discrete velocities. So, theoretically, the computational cost of the D2V30 model is twice that of the D2V15 model.
Compared with the standard multiphase lattice Boltzmann model, such as the well-known Shan-Chen D2Q9 model \citep{wang2009simulating},
the computational cost of D2V30 model  is about 4.4 times that of the D2Q9 model, greater than the ratio of the number of discrete velocities of the two DVMs $30/9$. This origins from  the more concise external force term of the Shan-Chen model.

(iv) The complexity of corresponding hydrodynamic equations increases
sharply with the degree of TNE, whereas that of the DBM does not have a
significant increase. We can formulate DBM with the desired order of
accuracy by incorporating more needed kinetic moments into $\mathbf{\Phi }%
_{n}$, and perform simulations without knowing the exact form of the
hydrodynamic equations.
For example, the D2V30 model satisfies 8 moment relations listed in Table \ref{table2} and describes
viscous stress and heat flux with third-order accuracy.
If one wants to achieve fourth-order accuracy, the physical model needs to further
satisfy higher-order moment relations $\mathbf{M}_{6}$ and $\mathbf{M}_{7,5}$.
Then the independent ones are $\mathbf{M}_{0}$, $\mathbf{M}_{1}$, $\mathbf{%
M}_{2}$, $\mathbf{M}_{3}$, $\mathbf{M}_{4}$, $\mathbf{M}_{5}$, $\mathbf{M}%
_{6}$ and $\mathbf{M}_{7,5}$. For a two-dimensional case, these moment
relations have 34 independent components, so a D2V34 model is sufficient to
discretize the phase space. Compared with the D2V30 model, the computational
cost of the D2V34 model is only increased by $13.33\%$.

(v) The Chapman-Enskog expansion acts as a bridge between the kinetic
description given by the DBM and continuum mechanics non-conserved
macroscopic dissipation and the conserved variables. Through it, we can
directly and strictly determine the minimal physical requirements, and
confirm the invariable moment sets $\mathbf{\Phi}_{n}$ for modeling a given
non-equilibrium state.
We also point out that, from the perspective of perturbation theory, conducting Chapman-Enskog
multi-scale analysis corresponds to imposing disturbance on the equilibrium system. The
convergence of Chapman-Enskog multi-scale expansion corresponds to the
system returning to its equilibrium state. If the disturbance is too strong,
such as due to the emergence of very steep gradients, leading to the
divergence of the Taylor expansion of the distribution function, then the
disturbance may trigger new structures or modes, and the Chapman-Enskog
analysis fails.
Fortunately, in the multiphase flow system, due to the existence of the
interface with a certain width, the gradient of the macroscopic quantity is
more gentle than that in the high-speed compressible single-phase system,
especially in the low-speed multiphase flow with large surface tension.

(vi) Strictly speaking, the discretization of continuous space inevitably brings non-physical symmetry breaking, and is a source of physical and numerical inaccuracies in the description of the system.
Coarse-grained physical modeling is inherently a trade-off.  The properties to be studied must be preserved, beyond scope of the given study can be sacrificed.
Of the many options available to meet the research needs of the current problem, the simplest or most economical is often the first choice.
The moment matching approach, shown in
\eqref{Moments}, affords the minimal discrete velocity sets for solving $\mathbf{\Phi }_{n}$, because the number of discrete velocities just equals the number of the independent kinetic moment relations. Therefore,
physically, DBM proposed in this way features the highest computational
efficiency. These outstanding advantages make DBM a particularly appealing
methodology for investigating non-equilibrium flows.
We can also build a more symmetrical DVM by increasing the number of discrete velocities. For example, we can construct a DVM through Hermite expansion~\citep{Shan-JFM-2006,Shan-PRE-2006,Shan-PRE-2019,Shan-JFM-2021,2008-Gan-PA} so that the DVM has better spatial symmetry at the cost of increasing the number of discrete velocities.
In this paper, we have chosen the simplest form matching the relevant
physical constraints for the problems at hand.

(vii)  For bulk flow far from the boundary, DBM includes three main pillars described in the Abstract. For near-wall flow, the mechanism for THNE may be greatly different from that for bulk flow, and consequently, a fourth pillar, construction of kinetic boundary conditions, is necessary.  A DBM for slip flow is referred to \citet{Zhang-AIPAdv-2022}.

(viii) The present work belongs to the category of single-component miscible multiphase
flow.
We have several works on DBM modeling and simulation of multi-component miscible
fluid flows~\citep{lin2016double,lin2017multi,zhang2020two,lin2021multiple}.
The extension of the present work to a multi-component immiscible fluid system is currently under consideration.
For multi-component immiscible fluid flows, self-aggregation still occurs between
different phases, and surface tension still exists between different
components. The core of DBM modeling for immiscible multiphase flow is the incorporation
of (a) intermolecular interaction in the same component and (b)
intermolecular interaction between different components.


\section{Numerical Simulations and Analysis}\label{sec3}

In this section, we first validate DBMs at various levels via three
benchmarks related to the equilibrium properties in bulk regions; then
evaluate carefully the multi-scale predictive capability of the DBMs by
investigating THNE features around the liquid-vapor interface; finally,
study the effects of relaxation time and surface tension coefficient on
thermal phase separation from the non-equilibrium  aspects. In all
simulations, we set $a=2$ and $b=0.4$, then the critical density and
temperature are $\rho ^{c}=1.30444$ and $T^{c}=1.88657$, respectively. To
improve the numerical stability and accuracy, the windowed fast Fourier transform
(FFT) scheme with $16$-th order in precision~\citep{2011-PRE-Gan-PS,2012-EPL-Gan-PS,2015-Gan-SM-PS} is utilized to discretize the spatial derivatives, and the second-order Runge-Kutta scheme is used to
calculate the temporal derivative. Notably, the periodic boundary condition
is inherently imposed in the $x$ and $y$ directions because of the
adoption of the FFT scheme.
For a system with non-periodic boundary conditions, the FFT scheme can also be applied
as long as the mirror operation is performed on the non-periodic boundary~\citep{gan2012physical}, at the cost of doubling the calculation amount.
All the parameters used here are dimensionless. Please refer to \cite{2019-Zhang-SM-PS} for
recovery of the actual physical quantities from the
numerical results.

\subsection{Verification and validation}\label{subsec3.1}

\subsubsection{Liquid-vapor coexistence curve}\label{subsubsec3.1.1}

\begin{figure}
	{\centerline{\epsfig{file=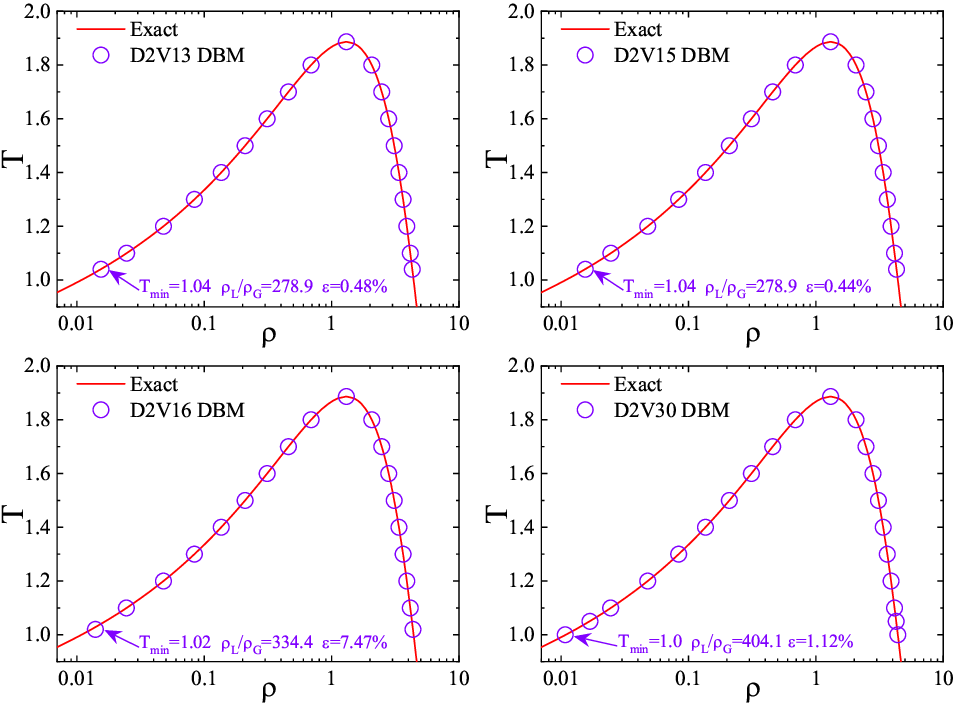,bbllx=104pt,bblly=282pt,bburx=579pt,bbury=623pt,
				width=0.9\textwidth,clip=}}}
	\caption{Comparisons of the coexisting densities predicted by various DBMs
		and Maxwell construction.}
\label{MC}
\end{figure}

Here the liquid-vapor coexistence curve test is used to check if the newly
constructed DBMs can reproduce the correct equilibrium thermodynamics of the
Carnahan-Starling system. For this problem considered, the initial
conditions are $(\rho, T, u_{x}, u_{y})=(\rho _{l}, 1.82, 0.0, 0.0)$, if
$N_{x}/4<x\leq 3N_{x}/4$; else $(\rho, T, u_{x}, u_{y})=(\rho
_{v}, 1.82, 0.0, 0.0) $, where $\rho _{l}=1.9643$ and $\rho _{v}=0.7569$ are the
theoretical liquid and vapor coexisting densities at $T=1.82$.
Parameters
are as follows $\tau =10^{-4}$, $K=1.5\times 10^{-4}$, $\Pr =0.01$, $%
N_{x}\times N_{y}=128\times 1$, $c=1.25$, $\Delta t=3\times 10^{-5}$, $%
\Delta x=\Delta y=4\times 10^{-3}$ for D2V13, D2V15 and D2V30 models. But in
the D2V16 model, the minimum spatial step that stabilizes the simulation is $%
\Delta x=\Delta y=5\times 10^{-3}$. The initial temperature drops by $0.01$
when a steady state has been reached, i.e., $|(u_{x}^{2}+u_{y}^{2})^{1/2}|_{%
\max }<10^{-6}$. Figure \ref{MC} displays the phase diagrams calculated from the
DBM simulations and Maxwell construction. In each panel, the two sets of
results are in good agreement with each other. However, the maximum density
ratio $R_{\max }=\rho _{\max }/\rho _{\min }$ and the lowest reduced
temperature $Tr_{\min }=T_{\min }/T^{C}$ that each model can undergo are
quite different.

The more the moment relations maintained by the DBM, the larger the maximum
density ratio and the lower the minimum reduced temperature that the model
can attain. Specifically, in the D2V13 model, $R_{\max }=278.9$ with the maximum relative error in density $\epsilon =|\rho _{\text{DBM}}-\rho _{\text{Exact}}|/\rho _{\text{Exact}}=0.48\%$ in the vapor branch ($\epsilon_{\text{max}}=0.28\%$ for the D2V30 model under the same density ratio). Whereas in
the D2V30 model, $R_{\max }$ can reach up to $404.1$ and $Tr_{\min }$ reduces
to $0.53$ with $\epsilon =1.12\%$ in the vapor branch.

There are several reasons that account for the fact that the incorporation of higher-order kinetic moments improves the density ratio that the model can sustain.
Physically, the incorporation of higher-order moment relation makes the
DBM model more approximate to the continuous Boltzmann-BGK equation with an
external force term, that is, the model is more physical; secondly, the
incorporation of higher-order moment relation makes DBM more accurate in
describing non-equilibrium effects, which greatly affects the computational
accuracy and stability, as shown in figure \ref{fig:Hydrodynamic quantities};
thirdly, the higher-order model is more capable of ensuring the positive entropy increase rate during the evolution process, with the entropy
growth rate defined as~\citep{2019-Zhang-SM-PS} $\frac{dS}{dt}=\int (\bm{\Delta }_{3,1}^{\ast }\cdot \bm{\nabla}\frac{1}{T}-%
\frac{1}{T}\bm{\Delta }_{2}^{\ast }{\colon }\bm{\nabla}\mathbf{u})d\mathbf{r}$;
fourth, the incorporation of higher-order moment relation is
equivalent to adopting more discrete velocities to discretize the phase
space. As the number of discrete velocity increases, the fluctuation of
distribution function $\delta f$ and the spatial gradient of distribution
function $\bm{\nabla}f$ decrease, which makes the higher-order model more
stable.
Lots of numerical examples suggest that the stability improvement of
higher-order model compared to lower-order ones is effective not only for
static (or quasi-static) tests, but also for dynamic cases.


\subsubsection{ Spurious velocity}\label{subsubsec3.1.2}

\begin{figure}
{\
\centerline{\epsfig{file=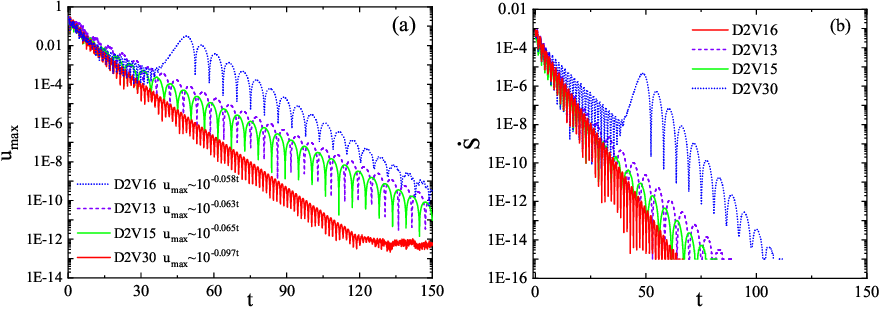,bbllx=115pt,bblly=378pt,bburx=545pt,bbury=528pt,
				width=0.99\textwidth,clip=}}}
\caption{
(a) Time evolutions of the maximum velocity $u_{\text{max}}$ calculated
from various DBMs. (b) Time evolutions of entropy increase rate.}
\label{us}
\end{figure}


Figure \ref{us} illustrates the time evolutions of the maximum velocity $%
u_{\max }$ for an initially non-equilibrium planar interface,
obtained from various DBM simulations performed using
the procedure and parameters exhibited in figure \ref{MC}, except for the initial temperature is
suddenly quenched from $1.45 $ to $1.10$, and the relaxation time $\tau
=2\times 10^{-4}$.
The spurious velocity, mainly located around
the liquid-vapor interface,
refers to the residual velocity which cannot be further reduced after a long
period of evolution~\citep{wagner2003origin}. Specifically, for the D2V30 model, spurious velocity
refers to the remaining current when $t>120$ in figure \ref{us}. The decay before $%
t<120$ is a physical process decided by the initial condition.

 Discarding the
early time transient regime and the very late time regime, where $u_{\max }$
stays almost unchanged, we find, $u_{\max}$ decreases with time $t$
approximately in the following way: $u_{\max }\sim 10^{-\gamma t}$. But the
exponent of the D2V30 model ($\gamma \approx 0.1$) is considerably larger
than that of other models ($\gamma \sim 0.061\pm 0.003$). When $t>120.0$, $%
u_{\max }$ decays to a negligible value $10^{-12}$ in D2V30 simulation under
the combined effects of NOMF and NOEF.
Figure \ref{us}(b) shows time evolutions of entropy increase rate for
the decaying process. When the system temperature is fixed, entropy increase
rate reduces to $\frac{dS}{dt}=\int -\frac{1}{T}\bm{\Delta }_{2}^{\ast }{%
\colon }\bm{\nabla}\mathbf{u}d\mathbf{r}$. The entropy increase rate
reflects the speed at which the system towards equilibrium, which is
qualitatively consistent with figure \ref{us}(a). The larger the entropy increase rate,
the faster the decay in $u_{\text{max}}$. Compared with lower-order models,
a faster decay in $u_{\text{max}}$ obtained from the D2V30 model is due to
a more accurate viscous stress the model recovers.
These findings suggest that the
retrieval of higher-order kinetic moments in physical modeling makes the DBM
more powerful in refraining spurious currents.


\subsubsection{Conservativeness of the new model}\label{subsubsec3.1.3}

\begin{figure}
{\
\centerline{\epsfig{file=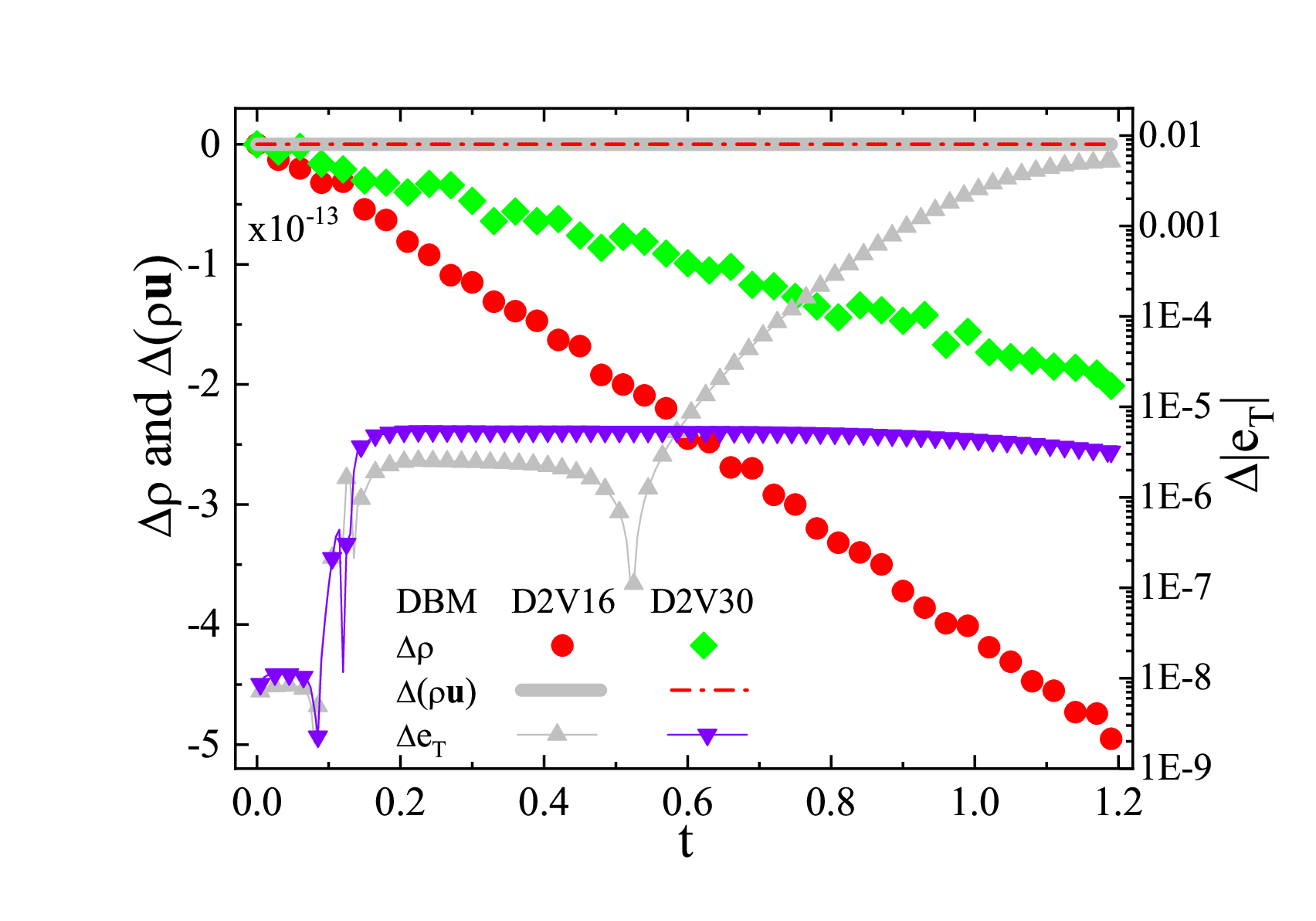,bbllx=9pt,bblly=5pt,bburx=830pt,bbury=541pt,
				width=0.7\textwidth,clip=}}}
\caption{Variations in $\rho $, $\rho \mathbf{u}$ and $e_{T}$
in a thermal phase separation process obtained from D2V16 and D2V30 DBMs.}
\label{conservation}
\end{figure}

To examine the conservativeness of the new model, we monitor variations in $\rho$, $\rho \mathbf{u}$ and $e_{T}$ in a thermal phase transition process
in figure \ref{conservation}. The initial conditions are $(\rho, T, u_{x}, u_{y})=(1.5+\Delta
, 1.0, 0.0, 0.0)$, where $\Delta $ is a random noise of the amplitude $0.001$
acting as heterogeneous nuclei and accelerating the kinetic process. The
remaining parameters are $N_{x}=N_{y}=128$, $\tau =10^{-3}$, $K=5\times
10^{-5}$, and $\Pr =0.1$. 
Even when the system is instantaneously quenched to a completely unstable
state, the two models maintain $\Delta \rho $ and $\Delta (\rho \mathbf{u})$
to machine accuracy.
Meanwhile, it is found that $\delta (\rho u)$ is two
orders of magnitude lower than $\delta (\rho )$. This is because: (a) $%
\delta (\rho u)=u\delta \rho +\rho \delta u$; (b) different from the
isothermal phase separation process, the release of latent heat causes the
changing of local temperature. The freedom in temperature results in a more
detailed and stricter local mechanical and thermodynamic equilibrium,
consequently hampers the full development of velocity, especially for the
case with large viscosity. The mean velocity $u_{\text{mean}}$ is
approximately on the order of $10^{-2}$ throughout the simulations we
considered. So $u\delta \rho \sim 10^{-15}$; (c) furthermore, it is
reasonable that variations in $\rho (t)$ and $\rho u(t)$ are proportional to
their fluctuations, $\delta \rho \propto (\rho _{\max }-\rho _{\min })\sim
o(1)$ and $\delta u\propto (u_{\max }-u_{\min })\sim o(10^{-2})$, so $\delta
u$ is two orders of magnitude lower than $\delta \rho $. Then $\rho \delta
u\sim 10^{-15}$.

In our simulations, the average energy density $e_{T}=\rho T-a\rho
^{2}+K|\bm{\nabla}\rho |^{2}/2+\rho u^{2}/2=-3.000015$.
Different from $\Delta \rho $ and $\Delta (\rho \mathbf{u})$, $\Delta e_{T}(t)$ oscillates and persistently
grows during the whole procedure mimicked by the D2V16 model.
Two reasons
account for the oscillation and growth in $\Delta e_{T}(t)$. First, the
appearance of numerous liquid-vapor interfaces in the spinodal decomposition
stage induces macroscopic quantity gradients and accumulated spatial
discretization errors. Second, the resulting macroscopic quantity gradients
serve as sources of HNE and TNE which may threaten the validity of the lower-order DBM. The maximum deviation of $%
e_{T}(t)$ is $0.01$ for the D2V16 model, while $O(10^{-6})$ for the D2V30 model,
demonstrating that the D2V30 model is more effective in maintaining energy
conservation and more capable of handling typical non-equilibrium  processes.
These variations are indicators of model accuracy, which significantly
affect the phase morphology, flow field distribution, and stability of simulation.

The results for two-dimensional lattices with 13, 15, 16, and 30 velocities
indicate that increasing the order of approximation of the lattice Boltzmann
equation enhances performance in maintaining thermodynamic consistency,
energy conservation, and refraining spurious currents.


\subsection{Multi-scale model and trans-scale ability}\label{subsec3.2}

In the DBM modeling, there exist two sets of macroscopic variables, the
conservative set ($\rho $, $\rho \mathbf{u}$, $e_{T}$) and the non-conservative
counterpart ($\bm{\Delta }_{m,n}$ and $\bm{\Delta }_{m,n}^{\ast}$,
$m-n \geq 2$), which vary in diverse time and length scales. Compared with the slowly varying conservative variables, the THNE quantities are quickly
varying non-conservative ones, with extremely short relaxation times on the
order of $10^{-10}$ second~\citep{Two-time-scales-POF-1999}.
In this subsection, we evaluate carefully the
multi-scale predictive capability of the DBMs for describing THNE features.
Specifically, we determine whether or not
the DBM can correctly capture the fast-changing THNE characteristics around an
evolving liquid-vapor interface. Indeed, understanding precisely the fine
structure and the specific status of non-equilibrium state between two phases
at coexistence is of central importance in a variety of fields, because
these non-equilibrium characteristics determine the transport efficiency of
mass and energy across an interface, as well as the growth dynamics and
morphological behaviors.

To date, significant progresses have been made by pioneers in the
non-equilibrium effects of multiphase flows.
Frezzotti \emph{et al}. comprehensively studied the rich array of non-equilibrium effects of
multiphase flows through numerically solving the Enskog-Vlasov equation by
DSMC (Enskog-Vlasov DSMC) and through the MD method \citep
{Frezzotti-POF-2005,Frezzotti-POF-2011,Frezzotti-AIP-2012,Frezzotti-PRF-2018,Frezzotti-POF-2019}%
, including the non-equilibrium structure of the vapor-liquid interface,
velocity slip at the liquid-vapor boundary, non-equilibrium evaporation, and
non-equilibrium stress, etc.
In addition to microscopic MD and mesoscopic DSMC, recently, the extended
macroscopic moment method, in particular, the regularized moment method \citep%
{2003-Struchtrup-R13,Struchtrup-Book,Torrilhon-ARFM-2016} derived as
approximations of the Boltzmann equation by means of the order-of-magnitude
method has been rapidly developed and extended to the field of multiphase
flows~\citep{2017-Struchtrup-multiphase-POF,2022-Struchtrup-multiphase-JFM}
by Struchtrup \emph{et al}.
In \cite{2017-Struchtrup-multiphase-POF}, the macroscopic evaporation
boundary conditions for regularized 13 (R13) moment equations have been
derived from microscopic interface conditions of the Boltzmann equation,
which extends the application range of the R13 equation to multiphase flows in
the transition regime. More recently, in \cite
{2022-Struchtrup-multiphase-JFM}, a set of 26 moment equations for the
Enskog-Vlasov equation has been derived by using the Grad moment method,
which provides a computationally efficient and functionally unified
approach for ideal and non-ideal fluid flows far from equilibrium. Besides,
the coupled constitutive relations theory was developed to capture the
remarkable rarefaction effects at small scales in the liquid-vapor phase
transition processes~\citep{2021-JFM-NCCR}.
Meanwhile, the rarefaction effects in a head-on collision of two
identical droplets are investigated via various lattice Boltzmann models
with different degrees of precision~\citep{chen2022rarefaction}.
It is found that the rarefaction effects enhance the conversion from free energy to kinetic energy
and accelerate droplet coalescence.
Present \emph{et al} presented an original and fundamental molecular
dynamics lattice gas (MDLG) approach connecting lattice Boltzmann methods to
physical reality~\citep{parsa2017lattice,2020-Wagner-PRL,czelusniak2020force,
parsa2021nonuniqueness,pachalieva2021connecting}. The key
implementation of MDLG is an embedded coarse-graining procedure, i.e.,
mapping an MD simulation onto an LB framework. It permits to study of any
system that can be simulated with MD, including the nonideal system with
significant non-equilibrium fluctuations~\citep{2020-Wagner-PRL}, since the collision
operator is informed by an underlying MD simulation. This approach can be
considered as an optimal LBM or a more realistic coarse-grained fluctuating
method.
We stress that these microscopic (MD), mesoscopic (DSMC, LBM and DBM proposed here), and macroscopic (high-order moment method, extended hydrodynamics or generalized hydrodynamics models) models/methods, with different starting points and different emphases, have their own strengths and insights, complement each other, can not replace each other. In addition, the developments of these methods have stages.

Although extensive experimental and theoretical studies have been designed
and developed in the past decades to address this issue~\citep{1979-Advances-In-Physics,1977-JCP-Structure,1986-Adv-Chem-Phys,2002-Onuki-Book,2004-Science-Exp,
2006-NJP-Exp,2014-JPCC-Exp,2018-CIS-Exp}, the early stage of the
liquid-vapor transition, i.e., the spinodal decomposition stage, where
non-equilibrium  effects are much more
pronounced than the latter domain growth stage, still needs further investigations
from the viewpoint of non-equilibrium  thermodynamics.
Early studies tend to focus on conserved quantities and slow variables, because these characteristics are relatively stable and easy to grasp~\citep{Succi2001Book,Yeomans-1995-PRL-SD,Gonella-1997-PRL,XGL-2003-PRE,GLS-model-PRE2007,2011-PRE-Gan-PS,2012-EPL-Gan-PS,Sofonea-2004-PRE,
Gonella-2010-PRE,Gonella-2014-PRE,Onuki-PRE-2007,2002-Onuki-Book}.  The study of these conserved quantities and slow variables provides a basis for the study of fast-changing behavior
which is particularly relevant to the early stage of spinodal decomposition, a
process still far from being completely understood.

The reason is that these fast varying and complex THNE quantities springing up in
the spinodal decomposition stage are difficult to
measure and analyze. Moreover, related experiments are costly and
time-consuming. Theoretical analysis is usually limited to cases with
numerous simplifying assumptions and generalizations, such as the local
equilibrium assumption,
given small deviations from the equilibrium state and linear relationship
between fluxes and forces. How to overcome these limitations is a core
problem in physical modeling~\citep{1996-AMM-TNE,2001-PNAS-TNE,2016-PRE-ETH,2021-JFM-NCCR}.

The definition of any non-equilibrium strength depends on the perspective of
the study. Complex systems need to be investigated from multiple
perspectives. If we look at the system from $N$ angles, there are $N$ kinds
of non-equilibrium strengths. Therefore, if the $N$ kinds of non-equilibrium strengths
are taken as components to introduce a non-equilibrium  strength vector, it
should be more accurate and specific.
Below we use the four-component vector $\mathbf{S}_{\text{THNE}} = (\bm{\Delta}_{2}, \bm{\Delta}_{3,1}, \bm{\Delta}_{3}, \bm{\Delta}_{4,2} )$, whose analytical expressions have been derived, to roughly describe the strength of non-equilibrium.

\subsubsection{Weak THNE case}\label{subsubsec3.2.1}

According to the analytical formulas listed in Appendix \ref{appendix2} and Section \ref{subsec2.1},
four explicit factors control the intensities and structures of THNE
manifestations, i.e., the relaxation time $\tau$, the fluid velocity $%
\mathbf{u}$, the gradient force and interparticle force caused by gradients
of macroscopic quantities (GMQs). GMQs are implicitly affected by the
initial state and the duration of evolution, for example, the initial density
distribution, the quenching temperature, the surface tension coefficient,
and the Prandtl number, etc. Therefore, by adjusting these aspects, we
can design scenarios with weak, moderate, and strong THNE strengths.

\begin{figure}
{\
\centerline{\epsfig{file=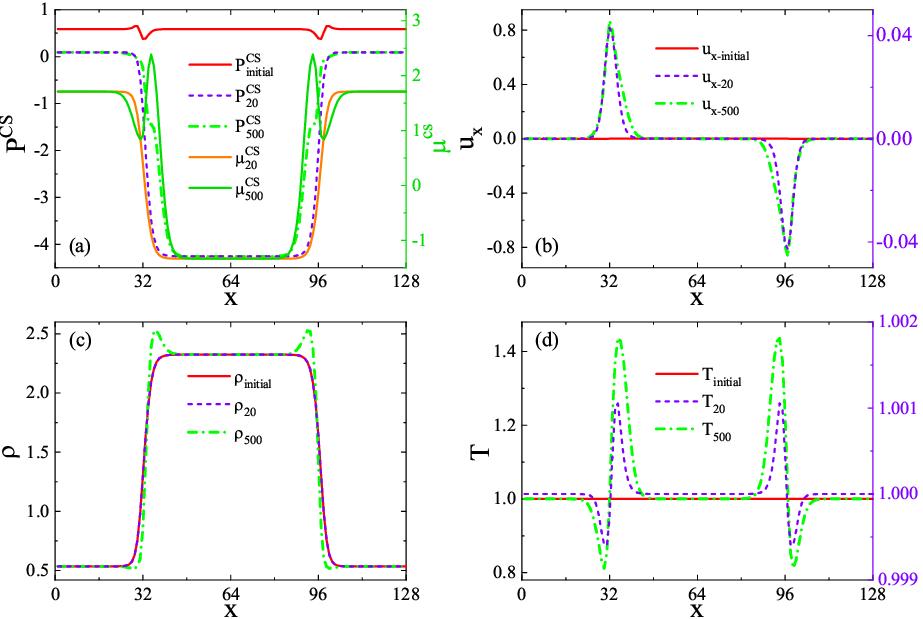,bbllx=108pt,bblly=301pt,bburx=545pt,bbury=600pt,
				width=0.99\textwidth,clip=}}}
\caption{Profiles of hydrodynamic quantities calculated from the D2V30 model
at three representative instants: $0$, $20$ and $500$ iterations.}
\label{profile}
\end{figure}

First, we investigate the weak THNE case. The initial state is the
equilibrium density profile $\rho _{\text{initial}}(x)$ at $T=1.74$, as
shown in panel (c) of figure \ref{profile}. The parameters are as follows $K=2.7\times
10^{-5}$, $\Delta t=2\times 10^{-5}$, $\Delta x=\Delta y=5\times 10^{-3}$,
others are consistent with those in figure \ref{MC}. When simulation starts, the system is
suddenly quenched to $T=1.0$. The profiles of
hydrodynamic quantities, including the chemical potential \citep{wagner2006thermodynamic,wen2017chemical,WEN-PRE-2020}
$\mu ^{\text{CS}}=RT[\frac{3-b\rho/4 }{(1-b\rho/4 )^{3}}+\ln \rho +1]-2a\rho -K\nabla
^{2}\rho$, 
calculated from the D2V30 model at
representative instants are shown in figure \ref{profile}, where $\phi _{n}$ represents
the macroscopic quantity after $n$ iterations. The following results have
been obtained: (i) the instantaneous decrease in temperature results in
the emergence of steep gradients in both pressure $P^{\text{CS}}$ and chemical potential $\mu ^{\text{CS}}$
near the liquid-vapor interfaces [see
panel (a)]; (ii) the pressure and chemical potential gradients drive the vapor phase across the
interface towards the liquid phase [see panels (b) and (c)], contributing to
the decrease (increase) in vapor (liquid) density, and the appearance of
remarkable velocity around the interface. After $20$ iterations, the maximum
velocity along the $x$ direction $u_{x\text{-max}}$ is approximately $0.04$, but
exceeds $0.80$ after $500$ iterations; (iii) with the development of phase
separation, latent heat is locally released to and absorbed by the
surrounding liquid and vapor phases, respectively. These processes lead to
violent changes in temperature and local mechanical imbalance. The maximum
temperature difference between the two phases $\Delta T_{\max }$ is
negligible at the very beginning of the procedure, but reaches $0.6T_{\text{%
initial}}$ at $t=0.01$ ($500$ iterations), which is the largest difference from an isothermal
case at a fixed temperature.

\begin{figure}
	{\
		\centerline{\epsfig{file=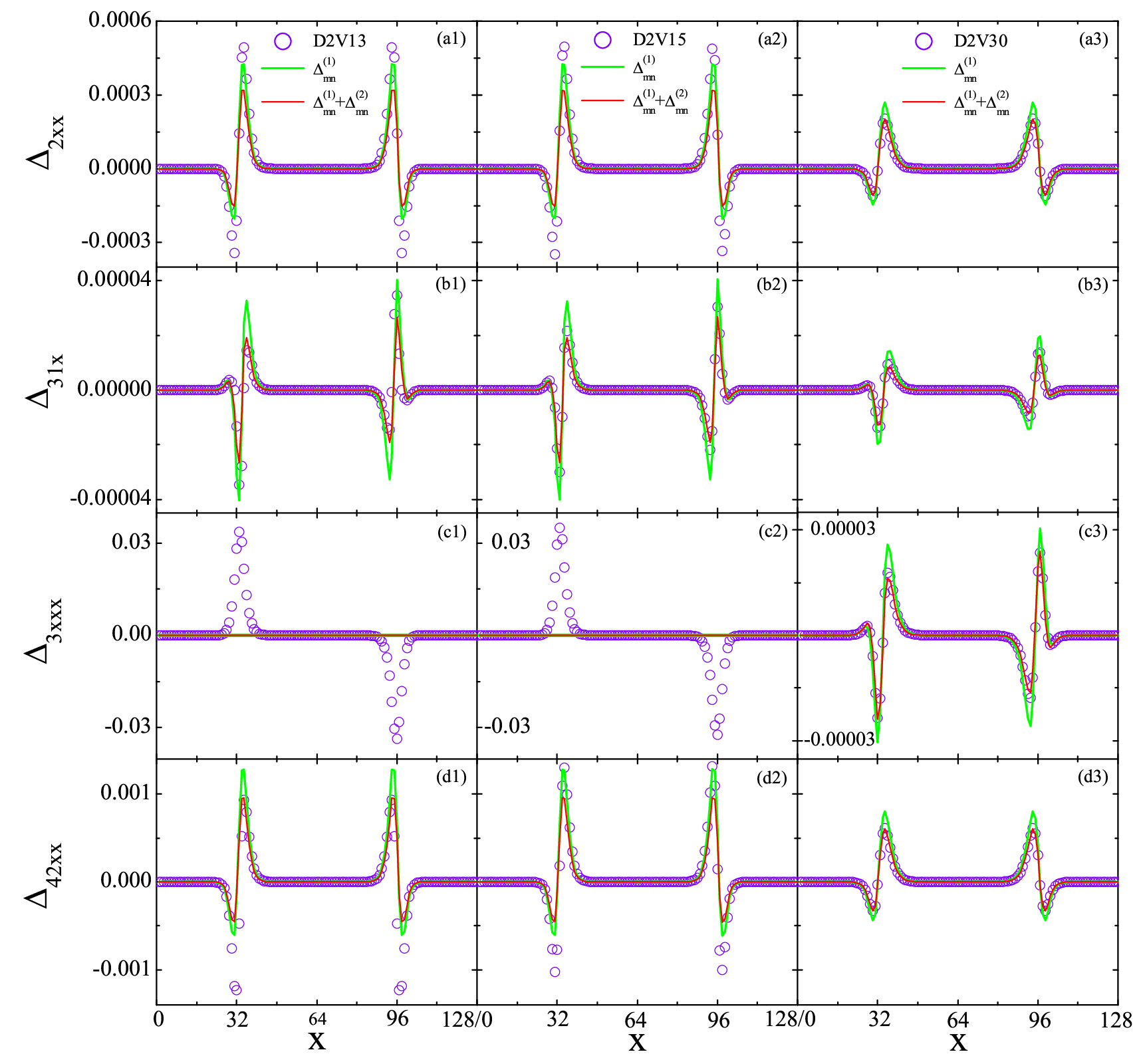,bbllx=15pt,bblly=5pt,bburx=840pt,bbury=790pt,
				width=0.99\textwidth,clip=}}}
	\caption{Weak case: THNE manifestations calculated from DBMs at various
		levels and theoretical analysis. Here $t=4\times 10^{-4}$ ($20$ iterations)
		and $\tau =10^{-4}$.}
\label{TNE-weak}
\end{figure}

Figure \ref{TNE-weak} portrays the specific THNE manifestations $\Delta _{2xx}$, $\Delta
_{3,1x}$, $\Delta _{3xxx}$, and $\Delta _{4,2xx}$ for figure \ref{profile} at $t=4\times
10^{-4}$ ($20$ iterations), where the D2V13 model at the NS level (left
column), the D2V15 model at the NS level (middle column), and the D2V30
model beyond the third-order super-Burnett level (right column) have been adopted. For
the convenience of comparisons, analytical solutions with the first and
second-order accuracies are plotted in each panel by solid lines. Figure \ref{TNE-weak}
shows the general characteristics when the system deviates from
thermo-hydrodynamic equilibrium: (i) THNE effects primarily concentrate in
the interfacial regions, where the interparticle force and gradient force dominate, and
disappear in the same manner as the GMQs vanish in the liquid and vapor bulk
phases. (ii) Owing to the limited action time of the gradient force and
interparticle force, THNE effects are not fully developed until this moment.
(iii) The first-order THNE $\bm{\Delta }_{m,n}^{(1)}\propto \tau $, is
always larger than the second-order THNE $\bm{\Delta }_{m,n}^{(2)}$ $%
\propto $ $\tau ^{2}$, demonstrating that $\bm{\Delta }_{m,n}^{(1)}$ is
the principal part of $\bm{\Delta }_{m,n}$ and, to some extent, the
effectiveness of the lower-order coarse-grained model. (iv) It is
interesting to find that $|\bm{\Delta }_{m,n}^{(1)}+\bm{\Delta }%
_{m,n}^{(2)}|$ is not larger than $|\bm{\Delta }_{m,n}^{(1)}|$, i.e.,
the second-order THNE invariably acts as negative feedback on the
first-order one, and weakens the total THNE intensity.

Beyond the aforementioned commonalities, the following distinctive
performances among various models should be thoroughly examined.
It is clear that, even with such a feeble THNE, the simulation
results of D2V13 and D2V15 models do not match well, either with the
first-order or the second-order theoretical solutions. More explicitly, the
two lower-order models tremendously overestimate the THNE level, especially
on the vapor side and for the $xxx$ component of $\bm{\Delta}_{3}$. The
discrepancies possess peak values approximately at the locations where the
gradient of the most associated physical quantity reaches its local maximum
(minimum).

Two key factors are responsible for the failure of the lower-order DBM in
capturing THNE accurately. On the one hand, to fully describe a given
non-equilibrium  quantity $\bm{\Delta }_{m,n}^{(j)}$, the number of the
required kinetic moments exactly retrieved by the discrete equilibrium
distribution function $f_{i}^{(0)}$ increases with increasing $m$, i.e., the rank
of $\bm{\Delta}_{m,n}$.
As shown in Table \ref{table1}, in the
formulation of $\bm{\Delta }_{3,1}^{(1)}$, $f_{i}^{(0)}$ should satisfy
the moment $\mathbf{M}_{5,1}$; when deriving $\bm{\Delta }_{4,2}^{(1)}$,
$f_{i}^{(0)}$ should satisfy higher-order moments $\mathbf{M}_{5,3}$ and $%
\mathbf{M}_{6,2}$. On the other hand, additional moment constraints on $%
f_{i}^{(0)}$ need to be selected and guaranteed with increasing $j$, i.e., the
order of accuracy of $\bm{\Delta }_{m,n}$. Concretely, a first-order
accuracy in the description of $\bm{\Delta }_{3}^{{}}$ can be achieved
by 
maintaining $\mathbf{M}_{4}$ and $\mathbf{M}_{5}$; while realizing a
second-order accuracy in the description of $\bm{\Delta }_{3}$, $%
f_{i}^{(0)}$ should further retain the additional thermodynamic one $\mathbf{%
M}_{6,4}$ ($\mathbf{M}_{5,3}$, in the 6th line of Table \ref{table1}, is not
independent of $\mathbf{M}_{5}$). Table \ref{table2} manifests that both the D2V13 and
D2V15 models cannot hold these necessary constraint relationships, even just
for the first-order accurate descriptions of $\bm{\Delta }_{3,1}^{(1)}$
and $\bm{\Delta }_{4,2}^{(1)}$. Among these THNE quantities, the
correct description of $\bm{\Delta }_{3}$ and $\bm{\Delta }_{4,2}$
requires the retrieval of the highest order moment relations. Thus, as
expected, the worst simulation results appear in panels (c1), c(2), d(1),
and d(2), provided by the two lower-order models.

Compared with the D2V13 model, the supplemental moment $\mathbf{M}_{5,1}$ which
is necessary for the formulation of $\bm{\Delta }_{3,1}^{(1)}$ and
maintained by the D2V15 model, plays a fairly negligible role in improving
the numerical precision of $\Delta _{3,1x}$, because the second-order
non-equilibrium  effects $\bm{\Delta }_{m,n}^{(2)}$ cannot be neglected in
contrast to the first-order $\bm{\Delta }_{m,n}^{(1)}$ during the early
stage of spinodal decomposition. Under this condition, considering the first-order THNE effects
only is not recommended.
Indeed, at the beginning of the phase separation, the gradients of
temperature and velocity approximately equal to zero. As a result, $\Delta
_{2xx}^{(1)}\simeq 0$, and $\Delta _{2xx}^{(2)}$ can be simplified to $\frac{%
T^{2}}{\rho }(\partial _{x}\rho )^{2}-T^{2}(\frac{\partial ^{2}\rho }{%
\partial ^{2}x})-\rho T^{2}\partial _{x}B_{x}$.
These higher-order terms in $\Delta _{2xx}^{(2)}$ do not appear in classical hydrodynamics, but is essential for modeling system far-from-equilibrium~\citep{Struchtrup-Book,2022-Struchtrup-multiphase-JFM}.
 Other non-equilibrium
measures can be analyzed similarly. That is, interparticle force and density
gradient force first trigger the second-order THNE, and afterward, stimulate the
first-order THNE through enlarging velocity and temperature gradients.
To evaluate under which circumstances the higher-order THNE should be taken
into account, a characteristic dimensionless parameter, called the relative
THNE strength $R_{\text{THNE}}=|\bm{\Delta }_{m,n}^{(j+1)}/\mathbf{%
\Delta }_{m,n}^{(j)}|$, may be introduced for the estimation of the relative
importance of the $(j+1)$-th order THNE effects to the $j$-th order one. The
$R_{\text{THNE}}$ for $\Delta _{2xx}$, $\Delta _{3,1x}$, $\Delta _{3xxx}$,
and $\Delta _{4,2xx}$ at the positions where these non-equilibrium  measures
own extreme values, are $0.250$, $0.349$, $0.212$, and $0.252$,
respectively. At earlier times, such as after $15$ iterations, $R_{\text{THNE}}$
is as high as $0.50$.
Therefore, higher-order DBMs with at least second-order accuracy for $%
\bm{\Delta }_{m,n}$, are undoubtedly needed, even for cases with faint
non-equilibrium  effects. Tables \ref{table1} and \ref{table2} indicate that the D2V30 model
satisfies all the needed kinetic moment relations. Consequently, the D2V30
results are consistent with the theoretical analysis at the second-order of accuracy.
Particularly, the remarkable deviations in panels (c1) and (c2) disappear in
panel (c3).


\subsubsection{Moderate THNE case}\label{subsubsec3.2.2}

\begin{figure}
	{%
		\centerline{\epsfig{file=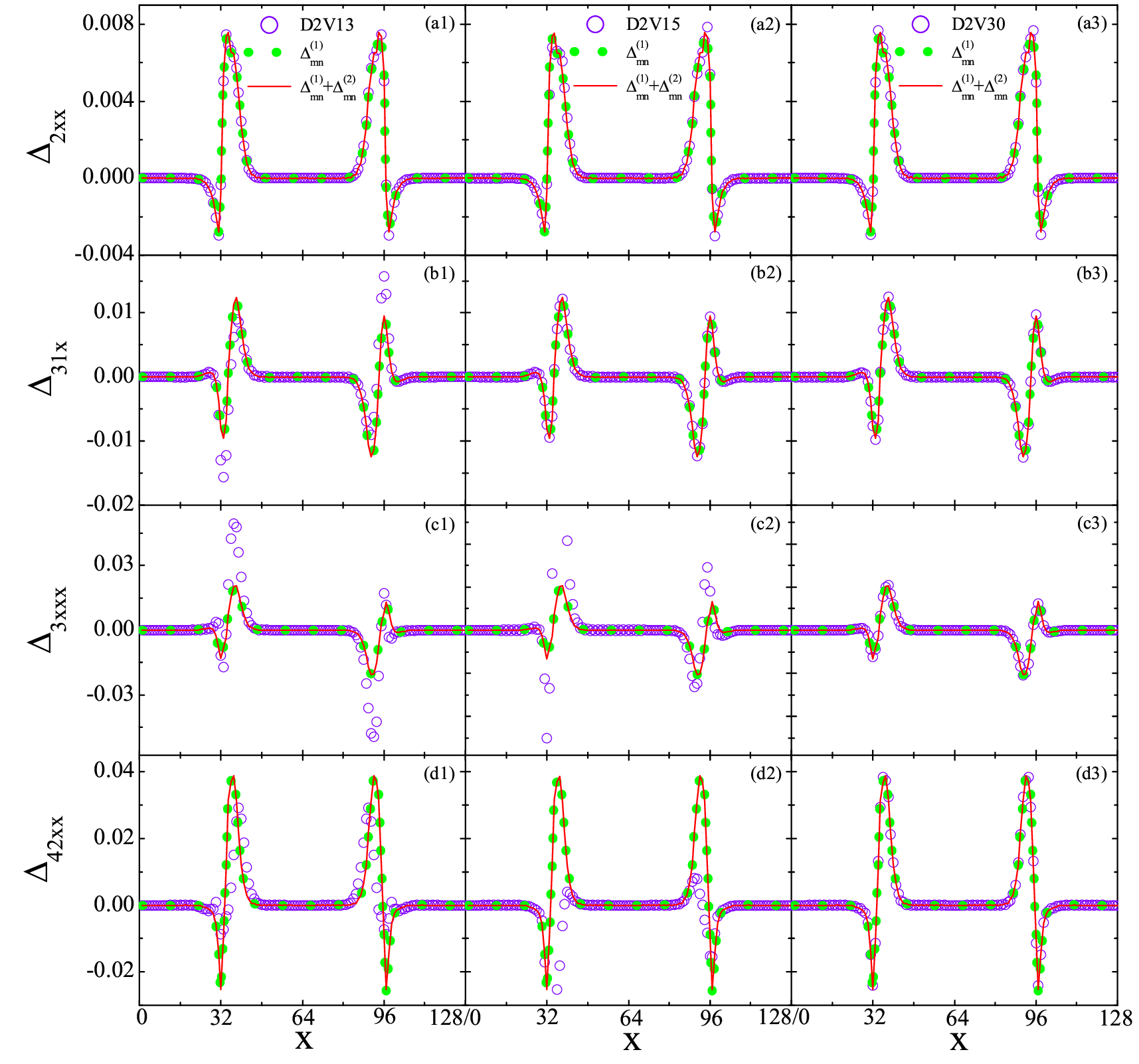,bbllx=15pt,bblly=5pt,bburx=840pt,bbury=790pt,
				width=0.99\textwidth,clip=}}}
	\caption{Moderate case: non-equilibrium  manifestations calculated from DBMs
		at various levels and theoretical analysis. Here $t=0.01$ ($500$ iterations)
		and $\tau =10^{-4}$.}
\label{TNE-moderate}
\end{figure}

As time evolves, the degree of phase separation becomes pronounced, with
extended phase-separating regions and prominent gradients of macroscopic
quantities (see dash-dot lines in figure \ref{profile}). The non-equilibrium  manifestations
at the same instant, exhibited in figure \ref{TNE-moderate}, indicate that the system enters
into the moderate THNE stage. Compared with figure \ref{TNE-weak}, the most distinctive
difference in figure \ref{TNE-moderate} is the enormous growth in $\bm{\Delta }_{m,n}$,
approximately $10\sim 1000$ times larger than those in the weak case.
For each kind of THNE quantity, the first-order analytical solution (denoted
by big green dots) overlaps with the second-order one (denoted by red solid
lines), illustrating that $\bm{\Delta }_{m,n}^{(2)}$ is negligible
compared with $\bm{\Delta }_{m,n}^{(1)}$. Correspondingly, the relative
THNE strength $R_{\text{THNE}}$ for $\Delta _{2xx}$, $\Delta _{3,1x}$, $%
\Delta _{3xxx}$, and $\Delta _{4,2xx}$ are as small as $0.006$, $0.002$, $%
0.018$, and $0.016$, respectively. For $\Delta _{2xx}$, good agreements
between simulations and theoretical solutions are depicted in panels (a1)
and (a2). This is understandable because either the D2V13 or D2V15 model is a
sufficient description of $\Delta _{2xx}^{(1)}$. Nevertheless, for $\Delta
_{3,1x}$, owing to the lack of indispensable kinetic moment $\mathbf{M}_{5,1}
$ in the D2V13 model, distinct discrepancies are found in panel (b1); but
are absent in panel (b2), obtained from the D2V15 model that meets
the constraint of $\mathbf{M}_{5,1}$. Similarly, for $\Delta _{3xxx}$ and $%
\Delta _{4,2xx}$, the D2V13 and D2V15 results mismatch the theoretical
profiles, because of the deficiencies in the necessary high order moments $%
\mathbf{M}_{4}$, $\mathbf{M}_{5}$, and $\mathbf{M}_{6,2}$. By contrast, as
shown in the third column of figure \ref{TNE-moderate}, the D2V30 results excellently coincide
with the theoretical results, regardless of the relative non-equilibrium
intensity. We also stress that, if we focus only on the hydrodynamic
quantities and constitutive relations (related to $\bm{\Delta }_{2}$ and
$\bm{\Delta }_{3,1}$), but not the evolution of the constitutive
relations (related to $\bm{\Delta }_{3}$ and $\bm{\Delta }_{4,2}$),
the D2V15 model is reliable and acceptable when $R_{\text{THNE}}$ is small
enough.


\subsubsection{Strong THNE case}\label{subsubsec3.2.3}

\begin{figure}
{\ \centerline{%
\epsfig{file=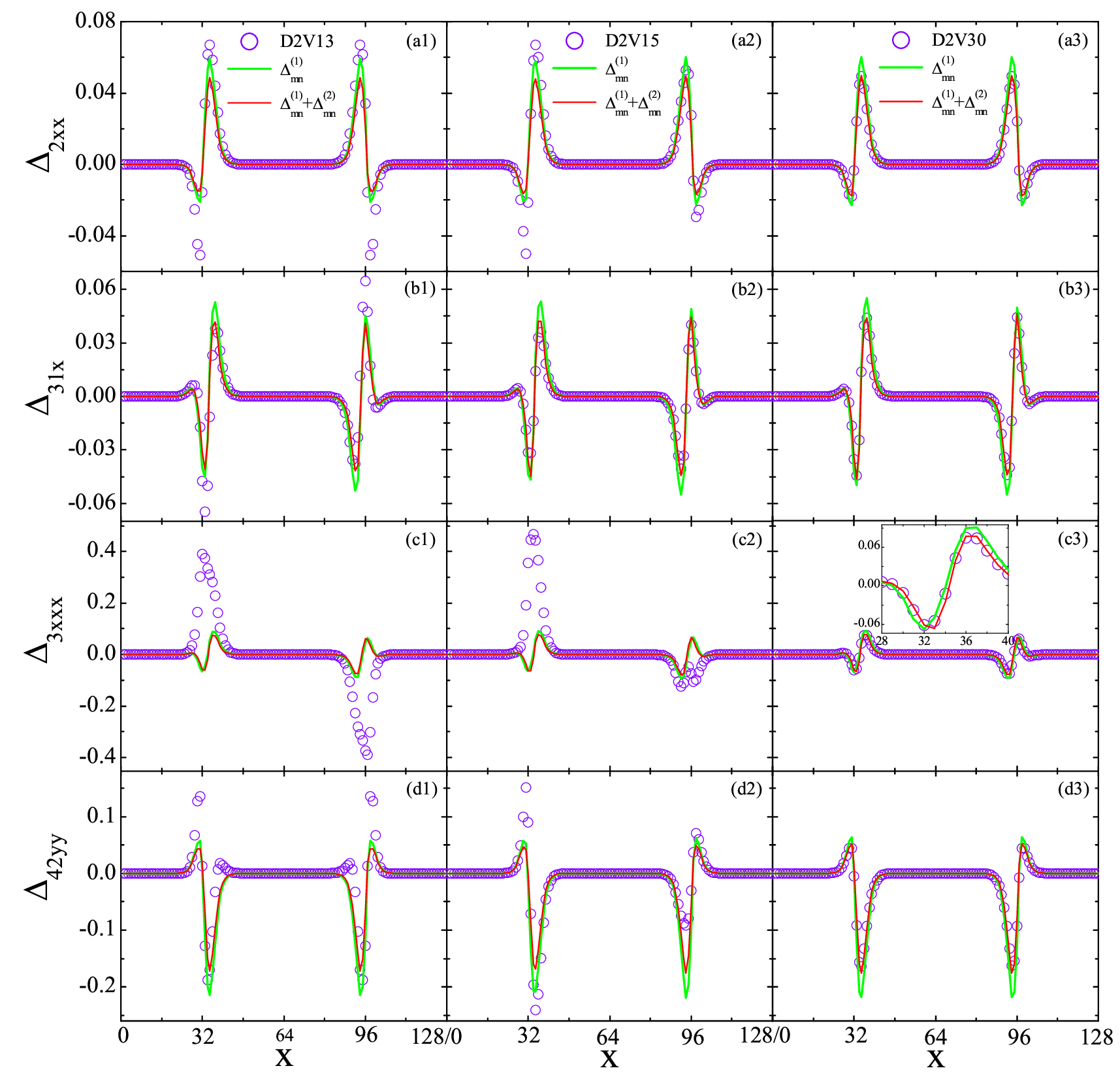,bbllx=13pt,bblly=5pt,bburx=840pt,bbury=796pt,width=0.99
\textwidth,clip=}}}
\caption{Strong case: non-equilibrium  manifestations
calculated from DBMs at various levels and theoretical analysis. Here $%
t=0.006$ ($300$ iterations) and $\tau =10^{-3}$.}
\label{TNE-strong2}
\end{figure}

To further examine the multi-scale ability of the models, we increase the
strength of THNE measures by straightforwardly increasing the relaxation time $\tau $ in figure \ref{TNE-strong2}.
In addition, another way to control the THNE strength is by adjusting the quenching depth $T/T_C$, which indirectly affects the non-equilibrium intensity
by changing the gradients of hydrodynamic quantities.
 With the increasing of THNE intensity, deviations between lower-order DBM simulations and theoretical
solutions become extraordinarily conspicuous (see the first and second
columns of figure \ref{TNE-strong2}). The relative THNE strength $R_{\text{THNE}}$ for $\Delta
_{2xx}$, $\Delta _{3,1x}$, $\Delta _{3xxx}$, and $\Delta _{4,2yy}$ rapidly
reach up to $0.213$, $0.248$, $0.169$, and $0.239$, respectively,
manifesting the importance of the second-order THNE and the
urgent need for a higher-order DBM. The D2V30 model meets the minimum
physical requirements in accurately describing the concerned THNE
quantities. Therefore, the reasonable agreements between D2V30 simulations
and corresponding theoretical solutions in the weak, moderate and strong
THNE cases comprehensively demonstrate that the D2V30 DBM is indeed a
multi-scale model and possesses better trans-scale ability than the D2V13
and D2V15 models.

To assess the application range of the DBM more completely, we resort to the
Knudsen number, a dimensionless parameter generally used to characterize
non-equilibrium or rarefaction degree of flows, and classify various flow
regimes. The Knudsen number $Kn=\lambda /L$, defined as the ratio of the
mean free path $\lambda =c_{s}\tau $ to a characteristic length we
focus on $L=\Theta /|\bm{\nabla}\Theta |$, where $c_{s}$ is the local speed
of sound, $\Theta $ a physical quantity. The maximum $Kn_{\max }$ calculated
by density for cases with weak, moderate, and strong THNE effects are $0.008$%
, $0.015$, and $0.141$, respectively, all beyond the application scope of NS
equations. Evidently, the applicable range of the D2V30 model has been extended into the early
transition regime.

\subsection{Effects of relaxation time on TNE}\label{subsec3.3}

\begin{figure}
{%
\centerline{\epsfig{file=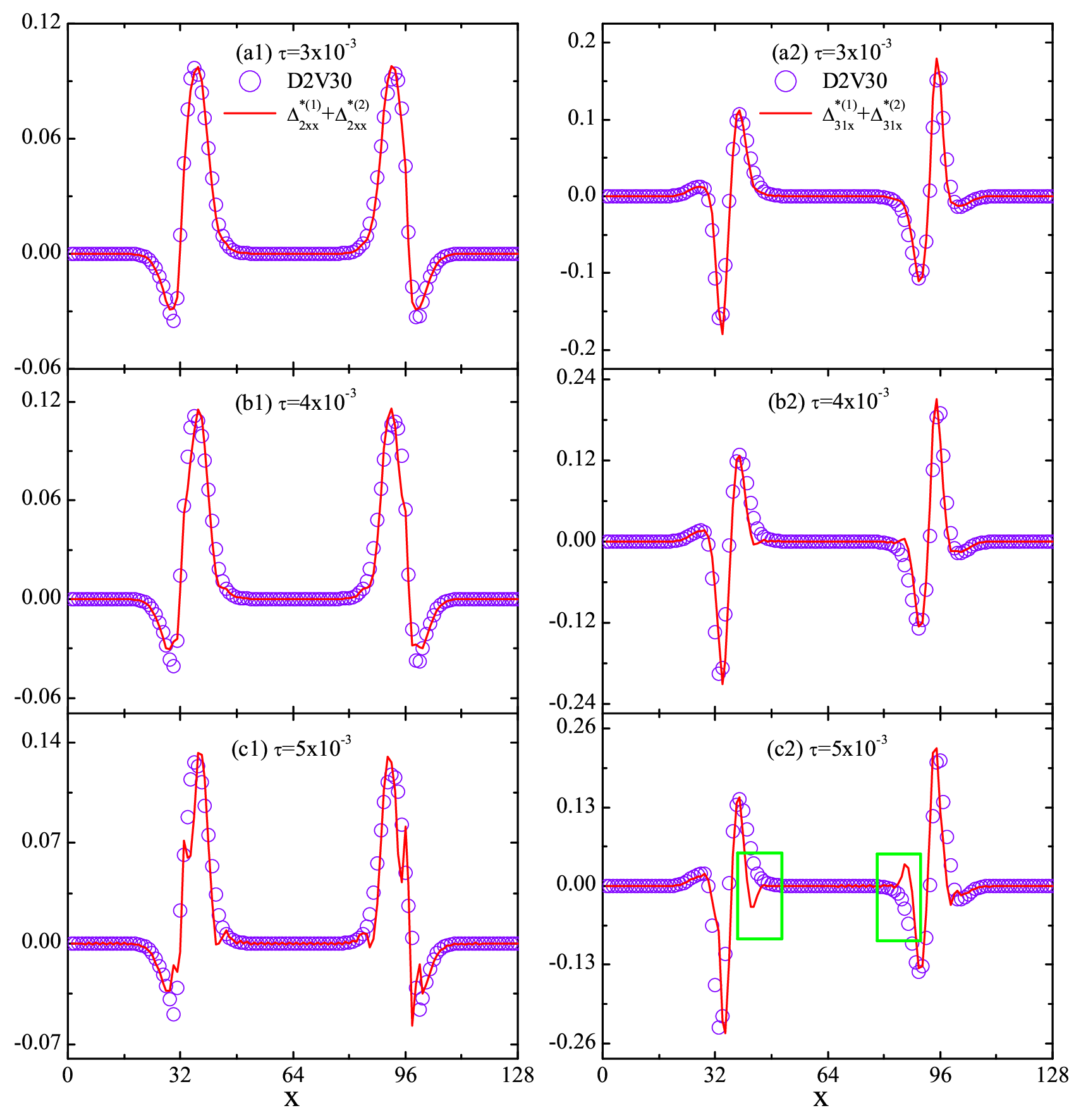,bbllx=0pt,bblly=7pt,bburx=816pt,bbury=855pt,
width=0.8\textwidth,clip=}}}
\caption{Effects of relaxation time on typical TNE manifestations: Viscous
stress (left column) and heat flux (right column). }
\label{TNE-tau}
\end{figure}

In this subsection, we investigate the multi-scale limit of the D2V30 model
through exploring whether the model can correctly reproduce higher-order
constitutive relations for multiphase flows over a wide range of relaxation
times and Knudsen numbers. Essentially, the accurate description of
constitutive relations is of fundamental importance to the simulation of
highly non-equilibrium  multiphase flows~\citep{Struchtrup-Book}. It is the constitutive relations
that directly determine the multi-scale ability and accuracy of a model. For
this purpose, in figure \ref{TNE-tau}, we exhibit the effects of relaxation time $\tau $
on the two typical TNE manifestations: viscous stress (left column) and heat
flux (right column) at $t=0.01$, where the initial conditions and parameters
in figure \ref{TNE-weak} are used, except for $K=5\times 10^{-5}$. To be seen is that,
the magnitudes of $\Delta _{2xx}^{\ast }$, $\Delta _{3,1x}^{\ast }$, and
the discrepancies between DBM simulations and the second-order theoretical
solutions, increase as $\tau $ increases. Specifically, for the case with $%
\tau =3\times 10^{-3}$, excellent coincidences can be observed between the two
counterparts; when $\tau $ increases to $4\times 10^{-3}$, good agreements
can be found; whereas when $\tau $ further increases to $5\times 10^{-3}$,
noticeable oscillations and distinctions, labeled by green rectangles,
emerge around the interfaces in panels (c1) and (c2), respectively.

Apart from the relative TNE intensity $R_{\text{TNE}}=|\bm{\Delta }%
_{m,n}^{\ast (j+1)}/\bm{\Delta }_{m,n}^{\ast (j)}|$, the aforementioned
TNE discrepancy between DBM simulation and theoretical solution, $\mathbf{%
\varkappa }_{\text{Dis}}=\bm{\Delta}^{\ast}_{\text{DBM}}-\bm{\Delta}^{\ast}_{\text{Exact}}$, or the relative TNE discrepancy $R_{\text{Dis}}=|(\bm{\Delta}^{\ast}_{\text{DBM}}-\bm{\Delta}^{\ast}_{\text{Exact}})/\bm{\Delta}^{\ast}_{\text{Exact}}|$ can be regarded as a non-equilibrium  criterion for assessing
the effectiveness of a coarse-grained model and appropriateness of the
coarse-graining process, as shown in Section \ref{sec2}. In real simulations, only
when $R_{\text{TNE}}$ or $R_{\text{Dis}}$ is negligibly small, the proposed
DBM is physically sufficient and appropriate. Otherwise, higher-order DBM
that considers higher-order TNE effects should be adopted. Thus, from this point
of view, the critical relaxation time for D2V30 model is $\tau _{C}=4\times
10^{-3}$ with respect to this standard test. When $\tau >\tau _{C}$, the
present model is ineffective and unacceptable, contributions of high order
departures from equilibrium distribution function $f^{(j)}$ ($j>2$) should
be considered in physical modeling.

\begin{figure}
{%
\centerline{\epsfig{file=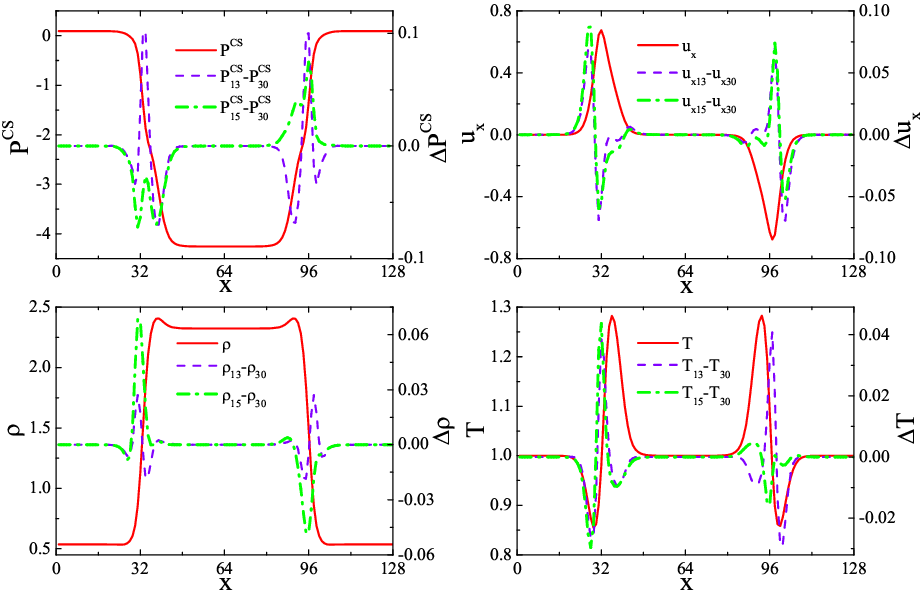,bbllx=112pt,bblly=305pt,bburx=561pt,bbury=603pt,
width=0.95\textwidth,clip=}}}
\caption{Hydrodynamic quantities calculated from the D2V30 model, and the
corresponding differences among D2V30, D2V15 and D2V13 models at $t=0.01$
when $\tau =4\times 10^{-3}$.}
\label{fig:Hydrodynamic quantities}
\end{figure}

Figure \ref{fig:Hydrodynamic quantities} further demonstrates that higher-order constitutive
relations are necessary for obtaining accurate hydrodynamic quantities and
fine structures of material and mechanical interfaces as the degree of non-equilibrium deepens ($\tau =4\times 10^{-3}$),
where $\varphi_{M}-\varphi _{N}$ represents hydrodynamic quantity difference calculated
from models with $M$ and $N$ discrete velocities.
 The maximum relative error in the velocity profile is $28.4\%$ at this moment, typically
located at the highly non-equilibrium  regime, but quickly
reaches up to $70.0\%$ after $200$ iterations with the accumulation of
errors. This kind of error, belonging to the physical modeling aspect,
cannot be reduced or eliminated by improving the accuracy of the algorithm.
The inaccuracy of D2V13 and D2V15 models is due to the crucial missing of
some necessary higher-order moments required for recovering $f^{(j)}$ ($%
j\geq 2$).

\begin{figure}
{%
\centerline{\epsfig{file=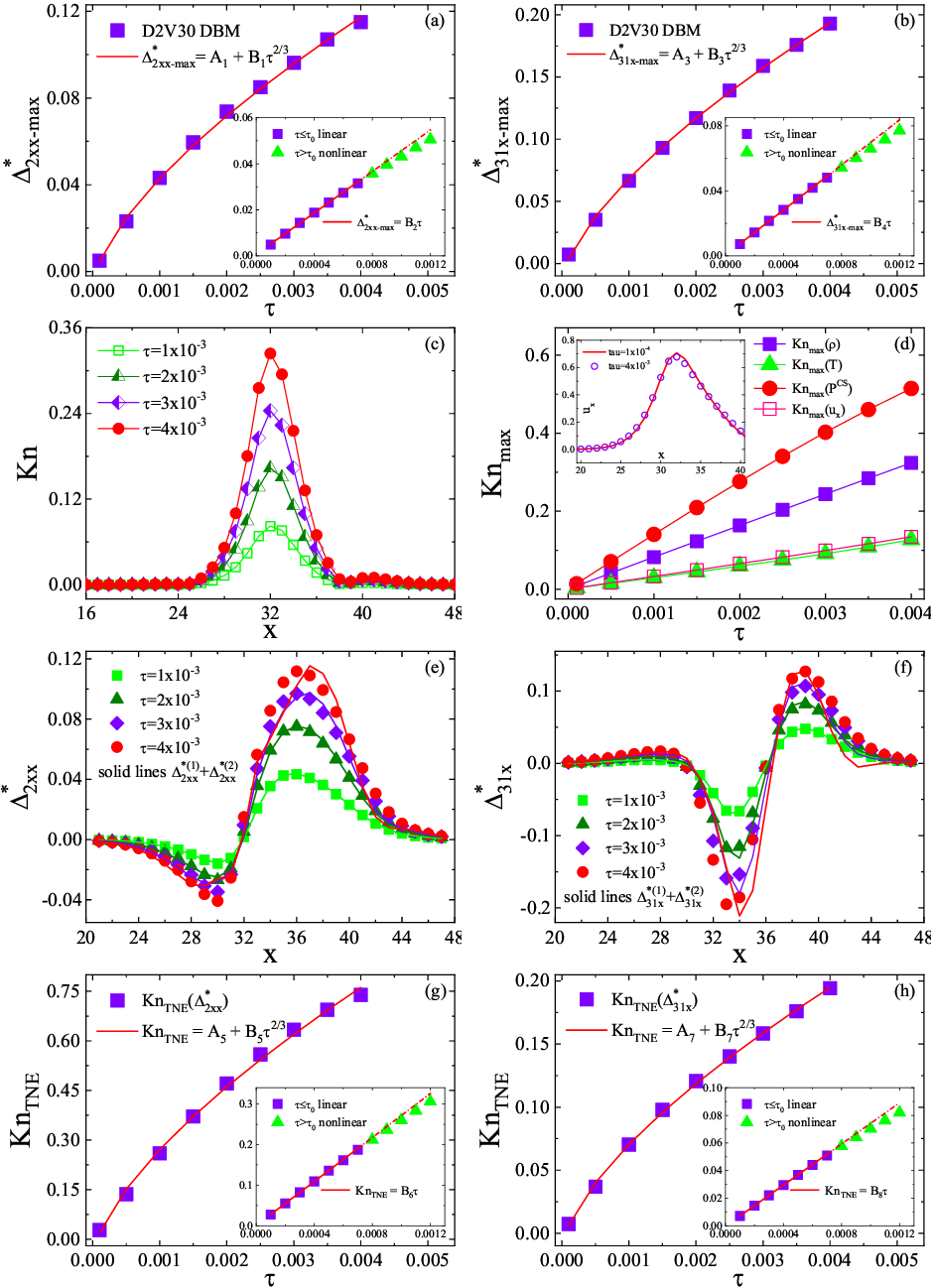,bbllx=85pt,bblly=80pt,bburx=565pt,bbury=708pt,		
width=0.96\textwidth,clip=}}}
\caption{Effects of relaxation time on the maximum viscous stress (a), the
maximum heat flux (b), distributions of the local Knudsen number (c), the
maximum Knudsen number (d), distributions of viscous stress (e),
distributions of heat flux (f), the Knudsen number calculated from viscous
stress (g), and the Knudsen number calculated from heat flux (h). }
\label{TNE-tau2}
\end{figure}

\begin{figure}
{\centerline{			%
\epsfig{file=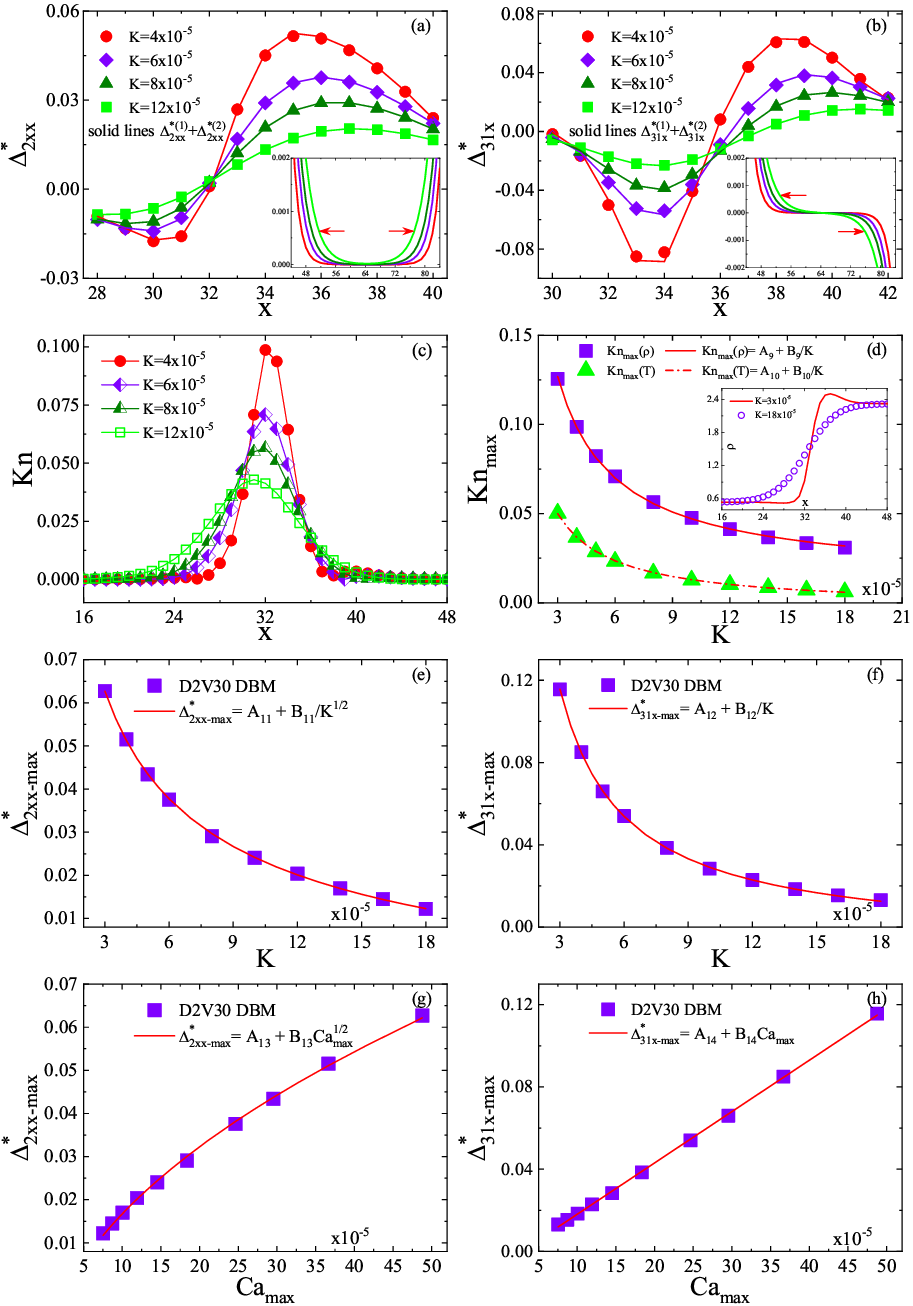,bbllx=105pt,bblly=80pt,bburx=549pt,bbury=709pt,width=0.96\textwidth,clip=}}%
}
\caption{Effects of surface tension coefficient on the distributions of
viscous stress (a), heat flux (b), and the local Knudsen number (c), the
maximum of Knudsen numbers calculated from density and temperature (d),
the maximum viscous stress (e) and the maximum heat flux (f). Relation between the maximum viscous stress and the maximum Capillary number (g), the relation between the maximum heat flux and the maximum Capillary number (h).}
\label{TNE-K}
\end{figure}

Subsequently, we illustrate the effects of relaxation time on the slowly and
quickly varying variables in figure \ref{TNE-tau2}. The first two panels in figure \ref{TNE-tau2}
demonstrates how relaxation time $\tau $ affects the maximum viscous stress $%
\Delta _{2xx\text{-max }}^{\ast }$ and the maximum heat flux $\Delta _{3,1x\text{-max }}^{\ast}$. Clearly, $\tau $ strengthens the intensities of $%
\Delta _{2xx\text{-max }}^{\ast }$ and $\Delta _{3,1x\text{-max }}^{\ast }$ in the
same manner. The relationship between $\Delta _{2xx\text{-max }}^{\ast }$ ($
\Delta _{3,1x\text{-max }}^{\ast }$) and $\tau $ can be divided into two stages:
linear and nonlinear. When $\tau \leq \tau _{0}$, $\Delta _{2xx\text{-max }%
}^{\ast }$ ($\Delta _{3,1x\text{-max }}^{\ast }$) increases linearly with $\tau $%
, $\Delta_{2xx{\text{-max}}}^{\ast}=B_{2}\tau $ ($\Delta _{3,1x{\text{-max }%
}}^{\ast }=B_{4}\tau )$; when $\tau >\tau _{0}$, the linearization is
no longer valid, the dependence of $\Delta _{2xx\text{-max }}^{\ast }$
($\Delta _{3,1x\text{-max }}^{\ast }$) on $\tau $ can be fitted by $\Delta _{2xx\text{-max }}^{\ast }=A_{1}+B_{1}\tau ^{2/3}$ ($\Delta _{3,1x\text{-max }}^{\ast
}=A_{3}+B_{3}\tau ^{2/3}$), manifesting the necessity of higher-order
constitutive relations for a system far-away-from equilibrium.
The fitting parameters are $A_1=-0.006$, $B_1=4.889$, $B_2=45.000$,
$A_3=-0.014$, $B_3=8.231$, $B_4=69.291$.
Of course,
the ending point of the linear constitutive relations provides a distinct
criterion to identify whether the system is near or far from equilibrium.
Panel (c) shows distributions of the local Knudsen number $Kn(x)$ with
various $\tau $. Similar to the behaviors of THNE, $Kn(x)$ becomes more
pronounced in the interfacial region and reach its maximum at the point at
which the gradient of quantity attains its peak value; and vanishes
exponentially in the bulk of the liquid and vapor regions. The maximum
interface Knudsen number $Kn_{\max }$ calculated from density exceeds $1/3$,
indicating that the D2V30 model is valid in the transition flow regime.
Different from the ideal gas system and the effects of surface tension, the
relaxation time $\tau $ only enlarges the amplitude of $Kn$, but does not
extend the non-equilibrium  region, or broaden the characteristic length scale
of a system, resulting in the approximately linear dependence of $Kn_{\max
}(\rho ,T,P^{cs},u_{x})$ on $\tau $, as shown in panel (d). This is because
the slowly varying variables appear to be slightly affected [for $u_{x}(x)$,
see the legend of figure \ref{TNE-tau2}(d)], or basically unaffected [for $\rho (x)$, $T(x)
$, $P^{CS}(x)$, not shown here] by relaxation time until this moment, but
noteworthily smoothed by the surface tension coefficient [see figures
\ref{TNE-K}(c)-(d)]. From this perspective, the interface Knudsen number calculated
from the slowly varying quantity cannot be viewed as a qualified 
parameter to characterize the non-equilibrium  behaviors at the early stage of
phase separation.

To solve this issue, we therefore resort to the quickly varying quantities,
i.e., the TNE measures, and Knudsen numbers calculated from them.
Figures \ref{TNE-tau2}(e)-(f) show effects of the relaxation time on viscous stresses and heat fluxes, where scatter symbols represent DBM simulation results, solid lines indicate the corresponding theoretical estimates.
As shown in panels (e) and (f),
the relaxation time substantially improves the
magnitudes of TNE measures and changes their structures. The characteristic
length scales are extraordinarily sensitive to relaxation time, position and
time. The last two panels suggest that the relaxation time significantly
enhances the maximum Knudsen numbers calculated from viscous stress $%
Kn_{\max }(\Delta _{2xx}^{\ast })$ (g) and heat flux $Kn_{\max }(\Delta
_{3,1x}^{\ast })$ (h). The $Kn_{\text{TNE}}\sim \tau $ relation behaves
qualitatively similar to the one between $\Delta _{2xx\text{-max }}^{\ast }$ ($%
\Delta _{3,1x\text{-max }}^{\ast }$) and $\tau $. The moderate growth in $Kn_{%
\text{TNE}}$ when $\tau >\tau_{0}$ results from the second-order TNE effect
that refrains the total TNE intensity.

\subsection{Effects of surface tension on TNE}\label{subsec3.4}

Here, the focus is on how and to what extent surface tension affects the
dynamic patterns and TNE features.
Figures \ref{TNE-K}(a)-(b) show the viscous stresses and heat fluxes at $t=0.01$ for cases with various surface tension
coefficients $K$, where symbols stand for simulation results, and solid lines in panels (a)-(b) and their legends represent the corresponding second-order
analytical solutions. The initial conditions and other parameters are the
same as those in figure \ref{TNE-strong2}. As shown, the surface tension effects are
threefold, refraining the local TNE intensity near the interface, but
expanding the TNE range and strengthening the TNE intensity away from the
interface (see the solid lines and the red arrows in legends for details),
through interface smoothing and extending.
This result can also be verified by figures \ref{TNE-K}(c)-(d), where the maximum of $Kn$ decreases but the region with non-zero $Kn$ increases with $K$, as a result of the capillary
wave broadening.
When $K$ varies in the interval $[3\times 10^{-5}$, $18\times 10^{-5}]$, the
dependence of $Kn_{\max }(\rho)$ and $Kn_{\max }(T)$ on $K$ can be fitted by $Kn_{\max }(\rho)=A_{9}+B_{9}/K$ and $Kn_{\max }(T)=A_{10}+B_{10}/K$, respectively, with $A_{9}=0.013$, $B_9=3.422\times 10^{-6}$,
$A_{10}=-0.003$, and $B_{10}=1.588\times 10^{-6}$.
The high sensitivity of density profile to surface tension coefficient [see the legend of figure \ref{TNE-K}(d)] is mainly due to the $\bm{\Lambda}$ term in \eqref{iki} which
is linearly proportional to $K$.

Accordingly, as plotted in figures \ref{TNE-K}(e)-(f), the maxima of $\Delta_{2xx}^{\ast}$ and
$\Delta_{3,1x}^{\ast}$ decrease synchronously with $K$ approximately in the following way, $\Delta_{2xx{\text{-max}}}^{\ast}=A_{11}+B_{11}/K^{1/2}$ and $\Delta_{3,1{\text{-max }} }^{\ast
}=A_{12}+B_{12}/K$, with $A_{11}=-0.023$, $B_{11}=4.667\times 10^{-4}$, $A_{12}=0.008$, and $B_{12}=3.711\times 10^{-6}$.
In other words, large resistances to mass and heat transfers appear at the interface, and increase with the surface tension coefficient.
The suppressive effect of surface tension on heat flux is much stronger than that on viscous stress at a small Prandtl number.
We further examine the effects of surface tension
on TNE in terms of the Capillary number $Ca$, which denotes the ratio between viscous force and interfacial tension,
$Ca=\mu u_{C}/\sigma$, with $u_{C}$ the characteristic velocity and $\sigma$ the interfacial tension between liquid and vapor phases.
For a planar interface, $\sigma$ can be specifically computed as
$\sigma =K\int_{-\infty }^{\infty }(\frac{\partial \rho }{\partial z})^{2}dz$.
Figures \ref{TNE-K}(g) and (h)
show relations among TNE measures and the maximum Capillary number $Ca_{\text{max}}$.
Clearly, $\Delta_{2xx{\text{-max}}}^{\ast }$ and $\Delta_{3,1x{\text{-max}}}^{\ast}$
increase with increasing $Ca_{\text{max}}$. The two
relations can be fitted by
$\Delta_{2xx{\text{-max} }}^{\ast}=A_{13}+B_{13}Ca_{\text{max}}^{1/2}$ and $\Delta_{3,1{\text{-max } }}^{\ast}=A_{14}+B_{14}Ca_{\text{max}}$, with fitting parameters $A_{13}=-0.021$, $B_{13}=3.761$, $A_{14}=-0.007$, and $B_{14}=249.513$.
The diametrically opposite trends in panels (e) and (g) [also panels (f) and (h)]
are due to the inverse relation between $Ca_{\text{max}}$ and $K$, $Ca_{\text{max}}\sim K^{-1}$.

\section{Conclusions and remarks}\label{sec4}

From the perspective of kinetic theory, the transport properties of a flowing
system are described by the distribution function $f$ or by the full sequence of its kinetic moments.
The conservative kinetic moments of $f$ include the density, momentum, and energy.
Acquisition of these conservative kinetic moments
is equivalent to knowing only the equilibrium distribution function $f^{(0)}$, namely perfect fluids free from dissipative effects.
How and to what extent the system deviates from the thermodynamic equilibrium remains completely unknown at this stage.

To know the distribution function $f$ is equivalent to knowing all the kinetic moments of $f$, which is neither possible nor necessary for the vast majority of situations.
Which kinetic moments need to be known for multi-scale modeling of complex flows
is the first key question that DBM endeavors to address. The second is how to detect, describe, present, and analyze THNE states.
Along this line, we present a framework for constructing multi-scale discrete Boltzmann
models for thermal multiphase flows ranging from continuum to transition flow regimes.
The DBM is a straightforward and efficient kinetic moment-matching method, whose basic
framework consists of three fundamental steps: (i)
establishment of the formal links with the modified continuous Boltzmann-BGK equation,
the extended hydrodynamic equations and the sought thermo-hydrodynamic
non-equilibrium  phenomena under consideration, (ii) formulation of the thermo-hydrodynamic
non-equilibrium  measures, and (iii) discretization of the particle velocity
space for the construction of the discrete equilibrium distribution function.
In addition, phase space manifolds spanned by the independent components
of non-conservative kinetic moments of $(f-f^{(0)})$ are introduced to describe
the corresponding non-equilibrium  states and effects.
The concept of metric distance from the origin in such phase space is extended to intuitively describe the non-equilibrium depth. The concept of metric distance between two points in such phase space is extended to intuitively describe the difference between two states.

In the case of the validity of Chapman-Enskog theory,  step (ii) permits
to select the minimum physical requirements for describing the
thermo-hydrodynamic non-equilibrium  quantities of interest to the desired order of
accuracy, that is, it determines the fewest required kinetic moments $\bm{\Phi
}_{n}=(\mathbf{M}_{0}$, $\mathbf{M}_{1}$, $\mathbf{...}$, $\mathbf{M}%
_{n})^{T}$ that the discrete equilibrium distribution function should retrieve
during the coarse-grained physical modeling process.
It also provides valuable byproducts: Higher-order constitutive relations and the most
relevant higher-order thermo-hydrodynamic non-equilibrium  manifestations
which are expected to improve the macroscopic modeling and
enhance the understanding of dynamic constitutive relations. 

The set of the fewest kinetic moments offers a unique perspective
to investigate the non-equilibrium  behavior of the
\textquotedblleft mesoscale\textquotedblright\ or \textquotedblleft
dilemma\textquotedblright\ situations where the continuum modeling fails and
 the molecular dynamics method lacks sufficient power to reach up to the scales of interest.
The research perspective and modeling accuracy depend on the specific elements
of moment sets, and can be adjusted and improved in time.
For instance, to access stronger non-equilibrium  effects, one needs to
include more higher-order moment relations into $\bm{\Phi }_{n}$, so that
the discrete Boltzmann model becomes more consistent with the continuous
Boltzmann equation far from equilibrium.
Importantly, the computational complexity of the DBM increases only
mildly in the process,  as opposed to the case of
generalized hydrodynamics, such as the Burnett or the super-Burnett
equations.

As model examples, two-dimensional discrete Boltzmann models with $13$, $15$
and $30$ discrete velocities at Navier-Stokes and beyond Burnett levels have
been proposed and evaluated. It is found that the incorporation of additional
higher-order kinetic moments helps to enhance the liquid-vapor
density ratio that the model can hold, to curb spurious currents
around the interface, and to ensure better mass, momentum, and energy
conservation in phase-separating systems.
The multi-scale predictive capability of the discrete Boltzmann model to
describe thermo-hydrodynamic non-equilibrium  features
increases with the number of
independent kinetic moments included by the model.
As fast varying variables, the
thermodynamic and thermo-hydrodynamic non-equilibrium  measures
provide more detailed information on the non-equilibrium multiphase
flow system than methods based on macroscopic conserved
quantities alone, thereby shedding new light on these complex states
of flowing matter far from equilibrium.
Future work includes
a more complete comparison among DBM results, high-order moment method~\citep{2017-Struchtrup-multiphase-POF,2022-Struchtrup-multiphase-JFM} and particle-based method \citep
{Frezzotti-POF-2005,Frezzotti-POF-2011,Frezzotti-AIP-2012,Frezzotti-PRF-2018,Frezzotti-POF-2019};
the incorporation of the external force in a more fundamental way, construction of appropriate kinetic boundary conditions \citep{Frezzotti-POF-2011,2017-Struchtrup-multiphase-POF,Zhang-AIPAdv-2022}, and consideration of the case with
faster phase separation, say the phase transition time scale is comparable to the molecular thermal relaxation.


\section*{Acknowledgements}
The authors sincerely thank the anonymous reviewers for
their valuable and instructive comments and suggestions, which were very
helpful for revising the manuscript. Also, we warmly thank
Dr. Tao Chen (Southern University of Science and Technology), Dr. Yudong Zhang (Zhengzhou University), Dr. Chuandong Lin (Sun Yat-sen University), Dr. Dejia Zhang (China University of Mining and Technology (Beijing)), and Dr. Bohai Chen (North China Institute of Aerospace Engineering) for
many helpful discussions.
We acknowledge support from the National Natural Science Foundation of China
(Grant Nos. 11875001 and 12172061), Natural Science Foundation of Hebei Province
(Grant Nos. A2021409001 and 226Z7601G), ``Three, Three and Three" Talent Project of Hebei
Province (Grant No. A202105005), Natural Science Foundation of Fujian Province
(Grant No. 2021J01652), CAEP Foundation (Grant No. CX2019033), the
opening project of State Key Laboratory of Explosion Science and Technology
(Beijing Institute of Technology) (Grant No. KFJJ21-16M) and Foundation of Laboratory of Computational Physics. SS gratefully acknowledges financial support from the European Research
Council under the Horizon 2020 Programme Grant Agreement n. 739964 (``COPMAT").

\section*{Declaration of interests}
The authors report no conflict of interest.


\appendix
\section{Determination of the relations among THNE measures and TNE measures}\label{appendix1}
According to the definitions in \eqref{THNE}-\eqref{TNE},
$\bm{\Delta}_{m,n}$ describes the combination effects of thermodynamic
non-equilibrium (TNE) and hydrodynamic non-equilibrium (HNE), which are
usually called the thermo-hydrodynamic non-equilibrium (THNE) effects; $%
\bm{\Delta }_{m,n}^{\ast }$ reflects molecular individualism on top of
organized collective motion, describing only the TNE effects. Next we give
out the relations among them:
\begin{eqnarray}
\bm{\Delta}_{2}^{\ast } &=&\mathbf{M}_{2}^{\ast }(f-f^{(0)})=\int_{-\infty
}^{\infty }(f-f^{(0)}){\mathbf{v}}^{\ast }{\mathbf{v}}^{\ast }d\mathbf{v} \\
&\mathbf{=}&\int_{-\infty }^{\infty }(f-f^{(0)})({\mathbf{v}}-{\mathbf{u}})({%
\mathbf{v}}-{\mathbf{u}})d\mathbf{v}  \notag \\
&\mathbf{=}&\int_{-\infty }^{\infty }(f-f^{(0)})({\mathbf{vv}}-{\mathbf{%
vu-uv+uu}})d\mathbf{v}  \notag \\
&=&\bm{\Delta}_{2}+\int_{-\infty }^{\infty }(f-f^{(0)})(-{\mathbf{vu}})d%
\mathbf{v+}\int_{-\infty }^{\infty }(f-f^{(0)})({\mathbf{-uv}})d\mathbf{v+}%
\int_{-\infty }^{\infty }(f-f^{(0)}){\mathbf{uu}}d\mathbf{v}  \notag \\
&=&\bm{\Delta}_{2}+[\int_{-\infty }^{\infty }(f-f^{(0)})(-{\mathbf{v}})d%
\mathbf{v]{u}}{\mathbf{-u[}}\int_{-\infty }^{\infty }(f-f^{(0)}){\mathbf{v}}d%
\mathbf{v]+[}\int_{-\infty }^{\infty }(f-f^{(0)})d\mathbf{v]}{\mathbf{uu}}.
\notag
\end{eqnarray}%
It is well known that $\int_{-\infty }^{\infty }(f-f^{(0)})d\mathbf{v=0}$
and $\int_{-\infty }^{\infty }(f-f^{(0)}){\mathbf{v}}d\mathbf{v=0}$, then
\begin{equation}
\bm{\Delta}_{2}^{\ast }=\bm{\Delta}_{2}\text{.}
\end{equation}%
\begin{eqnarray}
\bm{\Delta}_{3,1}^{\ast } &=&\mathbf{M}_{3,1}^{\ast
}(f-f^{(0)})=\int_{-\infty }^{\infty }(f-f^{(0)}){\mathbf{v}}^{\ast }\frac{{%
\mathbf{v}}^{\ast }\cdot {\mathbf{v}}^{\ast }}{2}d\mathbf{v} \\
&=&\int_{-\infty }^{\infty }(f-f^{(0)})({\mathbf{v}}-{\mathbf{u}})\frac{%
v^{2}-2{\mathbf{v}}\cdot {\mathbf{u}}+u^{2}}{2}d\mathbf{v}  \notag \\
&=&\int_{-\infty }^{\infty }(f-f^{(0)}){\mathbf{v}}\frac{v^{2}-2{\mathbf{v}}%
\cdot {\mathbf{u}}+u^{2}}{2}d\mathbf{v-}\int_{-\infty }^{\infty }(f-f^{(0)}){%
\mathbf{u}}\frac{v^{2}-2{\mathbf{v}}\cdot {\mathbf{u}}+u^{2}}{2}d\mathbf{v}.
\notag
\end{eqnarray}%
It is clear that $\int_{-\infty }^{\infty }(f-f^{(0)})\frac{v^{2}}{2}d%
\mathbf{v=0}$, hence
\begin{eqnarray}
\bm{\Delta}_{3,1}^{\ast } &=&\int_{-\infty }^{\infty }(f-f^{(0)}){\mathbf{v}}%
(\frac{v^{2}}{2}-{\mathbf{v}}\cdot {\mathbf{u}}+\frac{u^{2}}{2})d\mathbf{v}
\\
&=&\bm{\Delta}_{3,1}+\int_{-\infty }^{\infty }(f-f^{(0)}){\mathbf{v}}(-{%
\mathbf{v}}\cdot {\mathbf{u}}+\frac{u^{2}}{2})d\mathbf{v}  \notag \\
&=&\bm{\Delta}_{3,1}-[\int_{-\infty }^{\infty }(f-f^{(0)}){\mathbf{vv}}d%
\mathbf{v]}\cdot {\mathbf{u}}  \notag \\
&=&\bm{\Delta}_{3,1}-\bm{\Delta}_{2}^{\ast }\cdot {\mathbf{u}}\text{.}
\notag
\end{eqnarray}%
Similarly,
\begin{eqnarray}
\bm{\Delta}_{3}^{\ast } &=&\mathbf{M}_{3}^{\ast }(f-f^{(0)})=\int_{-\infty
}^{\infty }(f-f^{(0)}){\mathbf{v}}^{\ast }{\mathbf{v}}^{\ast }{\mathbf{v}}%
^{\ast }d\mathbf{v} \\
&\mathbf{=}&\int_{-\infty }^{\infty }(f-f^{(0)})({\mathbf{v}}-{\mathbf{u}})({%
\mathbf{v}}-{\mathbf{u}})({\mathbf{v}}-{\mathbf{u}})d\mathbf{v}  \notag \\
&=&\int_{-\infty }^{\infty }(f-f^{(0)})({\mathbf{vv}}-{\mathbf{vu}}-{\mathbf{%
uv}}+{\mathbf{uu}})({\mathbf{v}}-{\mathbf{u}})d\mathbf{v}  \notag \\
&=&\int_{-\infty }^{\infty }(f-f^{(0)})({\mathbf{vvv}}-{\mathbf{vuv}}-{%
\mathbf{uvv}}+{\mathbf{uuv-vvu+vuu+u{v}u-uuu}})d\mathbf{v}  \notag \\
&=&\int_{-\infty }^{\infty }(f-f^{(0)})({\mathbf{vvv}}-{\mathbf{vuv}}-{%
\mathbf{uvv-vvu}})d\mathbf{v}  \notag \\
&=&\bm{\Delta}_{3}-\int_{-\infty }^{\infty }(f-f^{(0)})({\mathbf{vuv}}+{%
\mathbf{uvv+vvu}})d\mathbf{v}  \notag \\
&=&\bm{\Delta}_{3}-\int_{-\infty }^{\infty }(f-f^{(0)}){\mathbf{vuv}}d%
\mathbf{v}-{\mathbf{u}}\bm{\Delta}_{2}^{\ast }-\bm{\Delta}_{2}^{\ast }{%
\mathbf{u}}  \notag \\
&=&\bm{\Delta}_{3}-(u_{\alpha }\Delta _{2\beta \gamma }^{\ast }+u_{\beta
}\Delta _{2\alpha \gamma }^{\ast }+u_{\gamma }\Delta _{2\alpha \beta }^{\ast
})\mathbf{e}_{\alpha }\mathbf{e}_{\beta }\mathbf{e}_{\gamma },  \notag
\end{eqnarray}%
\begin{eqnarray}
\bm{\Delta}_{4,2}^{\ast } &=&\mathbf{M}_{4,2}^{\ast
}(f-f^{(0)})=\int_{-\infty }^{\infty }(f-f^{(0)}){\mathbf{v}}^{\ast }{%
\mathbf{v}}^{\ast }\frac{{\mathbf{v}}^{\ast }\cdot {\mathbf{v}}^{\ast }}{2}d%
\mathbf{v} \\
&\mathbf{=}&\int_{-\infty }^{\infty }(f-f^{(0)})({\mathbf{v}}-{\mathbf{u}})({%
\mathbf{v}}-{\mathbf{u}})(\frac{v^{2}}{2}-{\mathbf{v}}\cdot {\mathbf{u}}+%
\frac{u^{2}}{2})d\mathbf{v}  \notag \\
&\mathbf{=}&\int_{-\infty }^{\infty }(f-f^{(0)})({\mathbf{vv}}-{\mathbf{%
vu-uv+uu}})(\frac{v^{2}}{2}-{\mathbf{v}}\cdot {\mathbf{u}}+\frac{u^{2}}{2})d%
\mathbf{v}  \notag \\
&=&\int_{-\infty }^{\infty }(f-f^{(0)})({\mathbf{vv}}-{\mathbf{vu-uv}})(%
\frac{v^{2}}{2}-{\mathbf{v}}\cdot {\mathbf{u}}+\frac{u^{2}}{2})d\mathbf{v}
\notag \\
&=&\int_{-\infty }^{\infty }(f-f^{(0)})({\mathbf{vv}}\frac{v^{2}}{2}-{%
\mathbf{vu}}\frac{v^{2}}{2}-{\mathbf{uv}}\frac{v^{2}}{2})d\mathbf{v}  \notag
\\
&&-\int_{-\infty }^{\infty }(f-f^{(0)})({\mathbf{vvv}}\cdot {\mathbf{u}}-{%
\mathbf{vu{\mathbf{v}}\cdot {\mathbf{u}}-uvv}}\cdot {\mathbf{u}})d\mathbf{v}%
+\int_{-\infty }^{\infty }(f-f^{(0)}){\mathbf{vv}}\frac{u^{2}}{2}d\mathbf{v}
\notag \\
&=&\bm{\Delta}_{4,2}-\bm{\Delta}_{3,1}{\mathbf{u-u}}\bm{\Delta}_{3,1}-%
\bm{\Delta}_{3}\cdot {\mathbf{u+}}\int_{-\infty }^{\infty }(f-f^{(0)})({%
\mathbf{vu{\mathbf{v}}\cdot {\mathbf{u}}}})d\mathbf{v}\text{ }{\mathbf{+}}%
\text{ }{\mathbf{u}}\bm{\Delta}_{2}\cdot {\mathbf{u+}}\frac{u^{2}}{2}%
\bm{\Delta}_{2}  \notag \\
&=&\bm{\Delta}_{4,2}-(\bm{\Delta}_{3,1}^{\ast }+\bm{\Delta}_{2}^{\ast }\cdot
{\mathbf{u}}){\mathbf{u-u(}}\bm{\Delta}_{3,1}^{\ast }+\bm{\Delta}_{2}^{\ast
}\cdot {\mathbf{u}})  \notag \\
&&-(\bm{\Delta}_{3}^{\ast }+\int_{-\infty }^{\infty }(f-f^{(0)}){\mathbf{vuv}%
}d\mathbf{v+}{\mathbf{u}}\bm{\Delta}_{2}^{\ast }+\bm{\Delta}_{2}^{\ast }{%
\mathbf{u}})\cdot {\mathbf{u}}  \notag \\
&&{\mathbf{+}}\int_{-\infty }^{\infty }(f-f^{(0)})({\mathbf{vu{\mathbf{v}}%
\cdot {\mathbf{u}}}})d\mathbf{v+}\text{ }{\mathbf{u}}\bm{\Delta}_{2}\cdot {%
\mathbf{u+}}\frac{u^{2}}{2}\bm{\Delta}_{2}  \notag \\
&=&\bm{\Delta}_{4,2}-\bm{\Delta}_{3,1}^{\ast }{\mathbf{u-u}}\bm{\Delta}%
_{3,1}^{\ast }-\bm{\Delta}_{3}^{\ast }\cdot {\mathbf{u}}-\bm{\Delta}%
_{2}^{\ast }\cdot {\mathbf{uu}}-\mathbf{u}\bm{\Delta}_{2}^{\ast }\cdot {%
\mathbf{u-}}\frac{u^{2}}{2}\bm{\Delta}_{2}^{\ast }.  \notag
\end{eqnarray}

According to the decomposition relations listed above, the formulas of THNE
measures ($\bm{\Delta}_{2}$, $\bm{\Delta}_{3,1}$, $\bm{\Delta}_{3}$, $%
\bm{\Delta}_{4,2} $) can be obtained only by knowing the expressions of TNE
measures ($\bm{\Delta}_{2}^*$, $\bm{\Delta}_{3,1}^*$, $\bm{\Delta}_{3}^*$, $%
\bm{\Delta}_{4,2}^*$).

	\begin{table}
\section{Expressions for the first- and second-order TNE measures}\label{appendix2}
		\begin{center}
		\begin{tabular}[b]{lc}
\hline
			$\Delta _{2\alpha \beta }^{\ast (1)}$ & $%
			\begin{array}{c}
				\Delta _{2xx}^{\ast (1)}=-\rho T\tau A_{1,-1} \\
				\Delta _{2xy}^{\ast (1)}=-\rho T\tau a_{1,1} \\
				\Delta _{2yy}^{\ast (1)}=-\Delta _{2xx}^{\ast (1)}%
			\end{array}%
			$ \\ \hline
			$\Delta _{3,1\alpha }^{\ast (1)}$ & $%
			\begin{array}{c}
				\Delta _{3,1x}^{\ast (1)}=-2\rho T\tau \partial _{x}T \\
				\Delta _{3,1y}^{\ast (1)}=-2\rho T\tau \partial _{y}T%
			\end{array}%
			$ \\ \hline
			$\Delta _{3\alpha \beta \gamma }^{\ast (1)}$ & $%
			\begin{array}{c}
				\Delta _{3xxx}^{\ast (1)}=-3\rho T\tau \partial _{x}T \\
				\Delta _{3xxy}^{\ast (1)}=-\rho T\tau \partial _{y}T \\
				\Delta _{3xyy}^{\ast (1)}=-\rho T\tau \partial _{x}T \\
				\Delta _{3yyy}^{\ast (1)}=-3\rho T\tau \partial _{y}T%
			\end{array}%
			$ \\ \hline
			$\Delta _{4,2\alpha \beta }^{\ast (1)}$ & $%
			\begin{array}{c}
				\Delta _{4,2xx}^{\ast (1)}=-3\rho T^{2}\tau A_{1,-1} \\
				\Delta _{4,2xy}^{\ast (1)}=-3\rho T^{2}\tau a_{1,1} \\
				\Delta _{4,2yy}^{\ast (1)}=-\Delta _{4,2xx}^{\ast (1)}%
			\end{array}%
			$ \\ 
			\hline 
			$\Delta _{2\alpha \beta }^{\ast (2)}$ & $%
			\begin{array}{c}
				\Delta _{2xx}^{\ast 2}/\tau ^{2}=-(2C_{1}A_{1,-1}+B_{1,-1})\rho
				T^{2}-[D_{1,-1}+\Gamma (u_{x})_{1,-1}+\Gamma (u_{y})_{1,-1}]\rho T \\
				-T^{2}F(\rho )_{1,-1}+\rho \Gamma (T)_{1,-1}+\frac{T^{2}}{\rho }\Gamma (\rho
				)_{1,-1} \\
				\Delta _{2xy}^{\ast 2}/\tau ^{2}=-\left( 2C_{1}a_{1,1}+b_{1,1}\right) \rho
				T^{2}-\left( d_{1,1}+2E_{1,1}\right) \rho T-2T^{2}\lambda (\rho )+2\rho
				\theta (T)+\frac{2T^{2}}{\rho }\theta (\rho ) \\
				\Delta _{2yy}^{\ast (2)}=-\Delta _{2xx}^{\ast (2)}%
			\end{array}%
			$ \\ \hline
			$\Delta _{3,1\alpha }^{\ast (2)}$ & $%
			\begin{array}{c}
				\Delta _{3,1x}^{\ast 2}/\tau ^{2}=-4\rho T^{3}\partial
				_{x}C_{1}-[A_{1,-1}B_{x}+a_{1,1}B_{y}+12C_{1}\partial
				_{x}T+F(u_{x})_{1,-1}+2\lambda (u_{y})]\rho T^{2} \\
				+\left( A_{1,-3}\partial _{x}T+a_{1,3}\partial _{y}T\right) 2\rho T \\
				\Delta _{3,1y}^{\ast 2}/\tau ^{2}=-4\rho T^{3}\partial
				_{y}C_{1}-[a_{1,1}B_{x}-A_{1,-1}B_{y}+12C_{1}\partial
				_{y}T-F(u_{y})_{1,-1}+2\lambda (u_{x})]\rho T^{2} \\
				+\left( a_{3,1}\partial _{x}T+A_{-3,1}\partial _{y}T\right) 2\rho T%
			\end{array}%
			$ \\ \hline
			$\Delta _{3\alpha \beta \gamma }^{\ast (2)}$ & $%
			\begin{array}{c}
				\Delta _{3xxx}^{\ast 2}/\tau ^{2}=-6\rho T^{3}\partial
				_{x}C_{1}-[A_{1,-1}B_{x}+6C_{1}\partial _{x}T+2\lambda (u_{y})]3\rho
				T^{2}+(A_{1,-2}\partial _{x}T+\partial _{y}u_{x}\partial _{y}T)6\rho T \\
				
				\Delta _{3yyy}^{\ast 2}/\tau ^{2}=-6\rho T^{3}\partial
				_{y}C_{1}-[A_{-1,1}B_{y}+6C_{1}\partial _{y}T+2\lambda (u_{x})]3\rho
				T^{2}+\left( A_{-2,1}\partial _{y}T+\partial _{x}u_{y}\partial _{x}T\right)
				6\rho T\\
				\Delta _{3xxy}^{\ast (2)}=2\Delta _{3,1y}^{\ast (2)}-\Delta _{3yyy}^{\ast
					(2)} \\
				\Delta _{3xyy}^{\ast (2)}=2\Delta _{3,1x}^{\ast (2)}-\Delta _{3xxx}^{\ast
					(2)}
			\end{array}%
			$ \\ \hline
			$\Delta _{4,2\alpha \beta }^{\ast (2)}$ & $%
			\begin{array}{c}
				\Delta _{4,2xx}^{\ast 2}/\tau ^{2}=-\left( 4C_{1}A_{1,-1}+B_{1,-1}\right)
				3\rho T^{3} \\
				-[D_{10,-2}+\Gamma (u_{x})_{1,-5}+\Gamma
				(u_{y})_{1,-5}+4e_{1,-1}-F(T)_{7,1}]\rho T^{2} \\
				+3\rho T\Gamma (T)_{7,1}-3T^{3}F(\rho )_{1,-1}+\frac{3T^{3}}{\rho }\Gamma
				(\rho )_{1,-1} \\
				\Delta _{4,2xy}^{\ast 2}/\tau ^{2}=-\left( 4C_{1}a_{11}+b_{1,1}\right) 3\rho
				T^{3}-[d_{1,1}+E_{1,1}-\lambda (T)]6\rho T^{2} \\
				+18\rho T\theta (T)-6T^{3}\lambda (\rho )+\frac{6T^{3}}{\rho }\theta (\rho )
				\\
				\Delta _{4,2yy}^{\ast 2}/\tau ^{2}=\text{ }\left(
				4C_{1}A_{1,-1}+B_{1,-1}\right) 3\rho T^{3} \\
				+[D_{2,-10}+\Gamma (u_{x})_{5,-1}+\Gamma
				(u_{y})_{5,-1}+4e_{-1,1}+F(T)_{1,7}]\rho T^{2} \\
				+3\rho T\Gamma (T)_{1,7}+3T^{3}F(\rho )_{1,-1}-\frac{3T^{3}}{\rho }\Gamma
				(\rho )_{1,-1}, \\
			\end{array}%
			$ \\ 
		\end{tabular}%
\end{center}
	\end{table}
\FloatBarrier
where ${A_{m,n}}=m\partial_{x}{u_{x}}+n\partial _{y}{u_{y}}$, ${a_{m,n}}%
=m\partial_{x}{u_{y}}+m\partial _{y}u_{x}$, ${B_{m,n}}=m\partial_{x}B_{x}+n%
\partial_{y}B_{y}$, ${b_{m,n}}=m\partial_{x}B_{y}+n\partial_{y}B_{x}$, ${%
D_{m,n}}=m{B_{x}}\partial_{x}T+n{B_{y}}\partial_{y}T$, ${d_{m,n}}=m{B_{y}}%
\partial_{x}T+n{B_{x}}\partial_{y}T$, ${E_{m,n}}=m\partial_{x}u_{x}%
\partial_{y}u_{x}+n\partial_{x}u_{y}\partial_{y}u_{y}$, ${e_{m,n}}%
=m\partial_{x}u_{x}\partial_{y}u_{y}+n\partial_{x}u_{y}\partial_{y}u_{x}$, $%
\theta(\cdot)=\partial_{x}(\cdot)\partial_{y}(\cdot)$, $\lambda(\cdot)=\frac{%
{{\partial^{2}}}}{{\partial y\partial x}}(\cdot)$, $F(\cdot){_{m,n}}=m\frac{{%
{\partial^{2}}}}{{\partial{x^{2}}}}(\cdot)+n\frac{{{\partial^{2}}}}{{\partial%
{y^{2}}}}(\cdot)$, $\Gamma(\cdot){_{m,n}}=m[\partial_{x}(\cdot)]{^{2}}%
+n[\partial_{y}(\cdot)]{^{2}}$, $C_1=C+C_q$.

\begin{table}
\section{Independent kinetic moments required for fully characterize the
first- and second-order TNE measures}\label{appendix3}
\begin{center}
\begin{tabular}[t]{lccc}
\hline
$\mathbf{M}_{m}$ & Independent kinetic moments \\ \hline
$\mathbf{M}_{0}$ & $%
\begin{array}{c}
M_{0}=\rho%
\end{array}%
$ \\ \hline
$\mathbf{M}_{1}$ & $%
\begin{array}{c}
M_{1x}=\rho u_{x} \\
M_{1y}=\rho u_{y}%
\end{array}%
$ \\ \hline
$\mathbf{M}_{2}$ & $%
\begin{array}{c}
M_{2xx}=\rho (T+u_{x}^{2}) \\
M_{2xy}=\rho u_{x}u_{y} \\
M_{2yy}=\rho (T+u_{y}^{2})%
\end{array}%
$ \\ \hline
$\mathbf{M}_{3}$ & $%
\begin{array}{c}
M_{3xxx}=\rho u_{x}(3T+u_{x}^{2}) \\
M_{3xxy}=\rho u_{y}(T+u_{x}^{2}) \\
M_{3xyy}=\rho u_{x}(T+u_{y}^{2}) \\
M_{3yyy}=\rho u_{y}(3T+u_{y}^{2})%
\end{array}%
$ \\ \hline
$\mathbf{M}_{4}$ & $%
\begin{array}{c}
M_{4xxxx}=\rho (3T^{2}+6Tu_{x}^{2}+u_{x}^{4}) \\
M_{4xxxy}=\rho u_{x}u_{y}(3T+u_{x}^{2}) \\
M_{4xxyy}=\rho (T+u_{x}^{2})(T+u_{y}^{2}) \\
M_{4xyyy}=\rho u_{x}u_{y}(3T+u_{y}^{2}) \\
M_{4yyyy}=\rho (3T^{2}+6Tu_{y}^{2}+u_{y}^{4})%
\end{array}%
$ \\ \hline
$\mathbf{M}_{5}$ & $%
\begin{array}{c}
M_{5xxxxx}=\rho u_{x}(15T^{2}+10Tu_{x}^{2}+u_{x}^{4}) \\
M_{5xxxxy}=\rho u_{y}(3T^{2}+6Tu_{x}^{2}+u_{x}^{4}) \\
M_{5xxxyy}=\rho u_{x}(3T+u_{x}^{2})(T+u_{y}^{2}) \\
M_{5xxyyy}=\rho u_{y}(3T+u_{y}^{2})(T+u_{x}^{2}) \\
M_{5xyyyy}=\rho u_{x}(3T^{2}+6Tu_{y}^{2}+u_{y}^{4}) \\
M_{5yyyyy}=\rho u_{y}(15T^{2}+10Tu_{y}^{2}+u_{y}^{4})%
\end{array}%
$ \\ \hline
$\mathbf{M}_{6,4}$ & $%
\begin{array}{c}
M_{6,4xxxx}=\frac{1}{2}\rho \lbrack
18T^{3}+3T^{2}(u^{2}+16u_{x}^{2})+2Tu_{x}^{2}(3u^{2}+5u_{x}^{2})+u^{2}u_{x}^{4}]
\\
M_{6,4xxxy}=\frac{1}{2}\rho
u_{x}u_{y}[24T^{2}+T(3u^{2}+10u_{x}^{2})+u^{2}u_{x}^{2}] \\
M_{6,4xxyy}=\frac{1}{2}\rho \lbrack
6T^{3}+Tu^{2}(9T+u^{2})+u_{x}^{2}u_{y}^{2}(10T+u^{2})] \\
M_{6,4xyyy}=\frac{1}{2}\rho
u_{x}u_{y}[24T^{2}+T(3u^{2}+10u_{y}^{2})+u^{2}u_{y}^{2}] \\
M_{6,4yyyy}=\frac{1}{2}\rho \lbrack
18T^{3}+3T^{2}(u^{2}+16u_{y}^{2})+2Tu_{y}^{2}(3u^{2}+5u_{y}^{2})+u^{2}u_{y}^{4}]%
\end{array}%
$ \\ \hline
$\mathbf{M}_{7,3}$ & $%
\begin{array}{c}
M_{7,3xxx}=\frac{1}{2}\rho
u_{x}[144T^{3}+16T^{2}(3u^{2}+5u_{x}^{2})+Tu^{2}(3u^{2}+20u_{x}^{2})+u^{4}u_{x}^{2}]
\\
M_{7,3xxy}=\frac{1}{2}\rho
u_{y}[48T^{3}+16T^{2}(u^{2}+5u_{x}^{2})+Tu^{2}(u^{2}+20u_{x}^{2})+u^{4}u_{x}^{2}]
\\
M_{7,3xyy}=\frac{1}{2}\rho
u_{x}[48T^{3}+16T^{2}(u^{2}+5u_{y}^{2})+Tu^{2}(u^{2}+20u_{y}^{2})+u^{4}u_{y}^{2}]
\\
M_{7,3yyy}=\frac{1}{2}\rho
u_{y}[144T^{3}+16T^{2}(3u^{2}+5u_{y}^{2})+Tu^{2}(3u^{2}+20u_{y}^{2})+u^{4}u_{y}^{2}]%
\end{array}%
$ 
\end{tabular}%
\end{center}
\end{table}
\FloatBarrier

\bibliographystyle{jfm}
\bibliography{jfm-instructions}
\end{document}